%% file: PFD_Micromech.tex
\documentclass[final,3p,number,sort&compress]{elsarticle}

\pdfoutput=1

\usepackage[english]{babel}
\usepackage[utf8]{inputenc}
\usepackage[T1]{fontenc}
\usepackage{charter}
\usepackage{setspace}

\usepackage{graphicx,calc}
\usepackage[export]{adjustbox}
\usepackage{color,xcolor}
\definecolor{grau}{HTML}{6F6F6F}

\usepackage{tikz}
\tikzstyle{line}=[draw, thick, -stealth]

\usepackage[font=footnotesize]{caption}
\usepackage[font=footnotesize]{subcaption}
\usepackage{float}
\usepackage{listliketab}

\usepackage{tabularx}
\usepackage{array}
\usepackage{multirow}
\usepackage{booktabs}

\usepackage{amsmath}
\usepackage{amssymb}
\usepackage{amsfonts}
\usepackage{amsxtra}
\usepackage{mathtools}
\usepackage{empheq}
\usepackage{stmaryrd}
\usepackage{icomma}
\usepackage{relsize}

\usepackage[colorlinks=false,hidelinks]{hyperref}

\renewcommand\fbox{\fcolorbox{white!0}{white!0}} 
\newcolumntype{Y}{>{\raggedright\arraybackslash}X}
\newcolumntype{L}[1]{>{\raggedright\let\newline\\\arraybackslash\hspace{0pt}}p{#1}}
\newcolumntype{P}[1]{>{\centering\arraybackslash}m{#1}}

\input{defs}
\journal{arXiv}

\begin{document}
	
\begin{frontmatter}
		
\title{Micromechanically motivated finite-strain phase-field fracture model to investigate damage in crosslinked elastomers}

\author[add1]{S. P. Josyula 
}\ead{siva.josyula@uni-saarland.de}
\author[add2]{ M. Brede  }
\author[add2]{O. Hesebeck }
\author[add2]{ K. Koschek  }
\author[add3]{ W. Possart }
\author[add2]{ A. Wulf  }
\author[add3]{   B. Zimmer  }
\author[add1]{S. Diebels 
}
\ead{s.diebels@mx.uni-saarland.de}
		
\address[add1]{Chair of Applied Mechanics, Saarland University, Saarbr\"ucken, Germany}
\address[add2]{Fraunhofer Institute for Manufacturing Technology and Advanced Materials, Wiener Stra{\ss}e 12, Bremen, Germany}
\address[add3]{Chair for Adhesion and Interphases in Polymers, Saarland University, Saarbr\"ucken, Germany}
\begin{abstract}
	A micromechanically motivated phase-field damage model is proposed to investigate the fracture behaviour in crosslinked polyurethane adhesive. The crosslinked polyurethane adhesive typically show viscoelastic behaviour with geometric nonlinearity. The finite-strain viscoelastic behaviour is modelled using a micromechanical network model considering shorter and longer chain length distribution. The micromechanical viscoelastic network model also consider the softening due to breakage/debonding of the short chains with increase in deformation. The micromechanical model is coupled with the phase-field damage model to investigate the crack initiation and propagation. Critical energy release rate is needed as a material property to solve phase-field equation. The energy release rate is formulated based on the polymer chain network. The numerical investigation is performed using finite element method. The force-displacement curves from the numerical analysis and experiments are compared to validate the proposed material model.
\end{abstract}
\begin{keyword}
	Crosslinked polymer network \sep Polyurethane adhesives \sep Fracture \sep Phase-field model \sep Micromechanical network model \sep Geometric nonlinearity.
\end{keyword}
\end{frontmatter}
\section{Introduction}
Crosslinked network polymer adhesives are becoming increasingly dominant in bonding technology for manufacturing various structural components in automobiles, shipbuilding and other areas of the modern manufacturing industry. These structural adhesives exhibit largely viscoelastic mechanical behaviour. Because of the resistance to fracture, these adhesives are particularly suitable for large differences in thermal expansion between different materials. 

Material models based on finite strain viscoelastic mechanical behaviour are modelled primarily for rubber materials \cite{HauptL}. A distinction is made between phenomenological material models formulated based on invariants or principal strains \cite{Mooney,Rivlin,Valanis,Ogden,Treloar,TRE46} and micromechanical network models, that consider the physics of the polymer network. Micromechanical network models for the crosslinked polymer materials consider the statistical distribution of the molecular chain network. For this purpose, the network is theoretically structured with interaction-free individual chains in the polymer network. From the statistical evaluation, it is possible to calculate the change in entropy of the chain as it is stretched, and it is possible to determine a force with the stretch \cite{KuhnW,KuhnG,Kuhn}.

In the transition from the single-chain model to the network model, the challenge is to assign a suitable stretching of the single-chain to the macro-level strain state. The affinity assumption that the three components of the chain length vector change at the same rate as the dimensions of the entire network seems obvious. However, a better description of actual material behaviour was obtained with the 8-chain model, which does not satisfy this condition, than with an affine network model with more accurate integration over the spatial directions \cite{ARR93,Boyce}. A systematic formulation of a non-affine network model, which includes the 8-chain model as a special case, was presented in \cite{MIE04}. The actual polymer network comprises shorter and longer chains, while simple network models consider only a constant average chain length. In \cite{SME99} the statistical molecular mass distribution was included in the model, which in turn is linked to the chain length distribution. Recently, an approach was developed to consider the chain length distribution following a model of polymerisation \cite{ITS16}. If the network model is extended to include internal variables, then dissipative processes such as the Mullins effect or permanent strains can be represented in addition to the elastic behaviour of the polymer. An isotropic description of the Mullins effect was obtained by formulating both the chain density and the chain length as a function of the largest stretch \cite{Marckmann}. 

Fracture is the complete or partial separation of a body, and a distinction is made between linear and non-linear fracture mechanics. A variety of failure hypotheses are proposed that include principal stresses, principal strains or comparative stress states. Griffith was the first to lay the foundation for fracture theory by introducing the energy required for crack propagation with the energetic fracture concept \cite{Griffith}. Models for both sharp and non-sharp crack discontinuities exist for the description of fracture mechanisms. Models describing sharp crack discontinuities include cohesive-zone models, but these models are computationally expensive and fail to investigate complex crack development. Computational models with crack discontinuities known as phase-field models allow the evaluation of crack propagation phenomena such as crack initiation or crack branching. Francfort and Marigo \cite{Francfort} proposed a variational approach based on the minimisation of the energy functional in the prediction of crack propagation phenomena based on continuum theory. Later, Bourdin et al. \cite{Bourdin} introduced a scalar-valued phase-field parameter distinguishing the damaged and intact material through regularising the variational formulation. With the minimisation principle, several contributions were made to investigate brittle \cite{Aranson,Eastgate,Corson,KarmaA,Hakim,KarmaK} and ductile fracture \cite{KuhnSM,KuhnCR,MieheAR} with phase-field theory. A combination of the micromechanical network model and phase-field model was proposed in \cite{MieheCS}. 

The current work focuses on formulating a phase-field damage model based on statistical chain mechanics to investigate fracture behaviour in crosslinked polyurethane adhesives. The mechanical behaviour is modelled using a statistically motivated finite-strain viscoelastic model, that considers the distribution of shorter and longer chains using the Wesslau distribution function \cite{wesslau}. The micromechanical model is coupled with a phase-field fracture with the crack energy release rate as the function of the distribution function. The proposed coupled formulation is implemented in deal. II and the parameters of the mechanical model are identified using a gradient-free algorithm for the curve fitting process programmed in MATLAB. Finally, the robustness of phase-field fracture for crosslinked polymers is investigated on the polyurethane adhesive based on DIN ISO 34-1 standards. 
\section{Constitutive model for network viscoelasticity}
The motivation of the work is to develop a micromechanically motivated approach to analyse fracture in crosslinked polymers like polyurethane adhesives. The mechanical behaviour considers the distribution of shorter and longer chain distribution in the material domain.
\subsection{Micromechanically motivated polymer network model}
The idea behind formulating a micromechanical free energy density based on the statistical distribution of polymer chains is achieved by multiplicatively extending isochoric free energy density ${\rm \Psi}(\rm I_1)$ with probable chain distribution function $g(\lambda_m)$. The probable chain distribution considers the debonding/breakage of the shorter chains with an increase in the applied load. The general form of the micromechanical free energy density is formulated as 
\begin{equation}
	W\left({\rm I_1}, \lambda_c\right) = {\rm \Psi}\left(\rm I_1 \right) \int \limits_{\lambda_c \left(\rm I_1 \right)}^{\infty}g(\lambda_m){\rm d}\lambda_m.
\end{equation}
The integral of $g(\lambda_m)$ is not possible because of the priori unknown $\infty$. Therefore, the free energy density is reformulated as  
\begin{equation}
	W\left({\rm I_1}, \lambda_c\right) = {\rm \Psi}\left(\rm I_1 \right) \left(\underbrace{\int \limits_{1}^{\infty} g(\lambda_m){\rm d}\lambda_m}_{\rm Term:1} - \underbrace{\int \limits_{1}^{\lambda_c \left(\rm I_1 \right)}g(\lambda_m){\rm d}\lambda_m}_{\rm Term:2}\right) = {\rm \Psi}\left(\rm I_1 \right) G(\lambda_{m}),
	\label{eq:micro1}
\end{equation}
where Term:1 corresponds to the probabilistic chain length distribution of the material at unstrained state\footnote{Probabilistic chain length distribution of the material at unstrained state is the sum of both inactive and active chains} and Term:2 corresponds to the probabilistic chain length distribution of the inactive chains\footnote{Inactive chains does not contribute to the strain energy density, since smaller chains do not transfer loads when deformation}. Probable chain stretch distribution $g(\lambda_m)$ is derived from the maximum chain stretch distribution ${\rm{w}(\lambda_m)}$. The distribution function ${\rm{w}(\lambda_m)}$ is formulated analogues to the Wesslau type molar mass distribution function by replacing molar mass with the current chain stretch $\lambda_m$ as:
\begin{equation}
	{\rm{w}(\lambda_m)} = \frac{1}{\beta \sqrt{\pi}}\frac{1}{(\lambda_m - 1)} {\rm{exp}} \left(\frac{-1}{\beta^2}\left({\rm{ln}}\left(\frac{(\lambda_m -1 )}{^0 \lambda_m}\right)\right)^2\right),
	\label{eq:lcs14}
\end{equation}
where the material constants polydispersity index $Q$ and average chain elongation $^M\lambda_m$, the parameters of ${\rm{w}(\lambda_m)}$ can be computed as follows
\begin{equation}
	\beta=\sqrt{2{\rm{ln}(Q)}};\hspace{3mm} ^N\lambda_m = \sfrac{^M\lambda_m}{Q};\hspace{3mm} ^0\lambda_m = \sqrt{^M\lambda_m \, ^N\lambda_m}.
\end{equation}
Based on the statistical chain mechanics of polymers, a random chain is formed of $N$ identical segments of length $l\, \rm mm$. the maximum length of chain $r_m$ and the random length of initial chain $r_0$ are calculated as
\begin{equation}
	r_m = Nl, \hspace{5mm} r_0=l\sqrt{N}.
\end{equation} 
and the current chain stretch is introduced as 
\begin{equation}
	r_c = \frac{1}{\sqrt{3}} l \sqrt{N} \sqrt{\rm I_1},
\end{equation}
where $\rm I_1$ is the first invariant of the left Cauchy-Green deformation tensor $\mathbf{B}$. The kinetics of chain terminology required to formulate the network model are current chain stretch $\lambda_c$ and maximum chain stretch $\lambda_m$. These quantities are introduced as 
\begin{equation}
	\lambda_c = \frac{r_c}{r_0} = \frac{\sqrt{\rm I_1}}{\sqrt{3}} \hspace{2 mm} \text{and} \hspace{2 mm} \lambda_m = \frac{r_m}{r_0}=\sqrt{N}.
\end{equation}
The current stretch always lies within the interval of minimum $\lambda_0$ and maximum $\lambda_m$ of chain stretch. The interval is interpreted as
\begin{equation}
	\lambda_0=1 \le \lambda_c = \frac{\sqrt{\rm I_1}}{\sqrt{3}} \le \lambda_{m}.
\end{equation}
Probable chain length distribution function $g(\lambda_m)$ is formulated from maximum chain stretch $\lambda_{m}$ and the chain stretch distribution ${\rm{w}}(\lambda_m)$
\begin{equation}
	\begin{aligned}
		g(\lambda_m) &= \frac{^N\lambda_m}{(\lambda_m-1)}{{\rm{w}}(\lambda_m)}\\
		&= a_0\frac{1}{(\lambda_m-1)^2} {\rm{exp}}\left(-a_1 \left( {\rm{ln}}\left(a_2(\lambda_m-1) \right) \right)^2 \right),
	\end{aligned}
	\label{eq:lcs18}
\end{equation}
with the parameters 
\begin{equation}
	a_0 = \frac{^N\lambda_m}{\beta\sqrt{\pi}};\hspace{3mm} a_1 = \frac{1}{\beta^2}; \hspace{3mm} a_2 = \frac{1}{^0\lambda_m}.
\end{equation}
The integral of Term:1 in equation \eqref{eq:micro1} for an unstrained state $\lambda_{m} =1$ is assumed as $g(\lambda_{m}=1)=1$ to avoid singularity. Indefinite integral of $g(\lambda_m)$ is dependent on $\lambda_m$ and is derived based on an error function $\rm erf(x)$ 
\begin{equation}
	G(\lambda_m) = \frac{a_0 a_2 {\rm{exp}}\left(\frac{1}{4a_1} \right)\sqrt{\pi}}{2\sqrt{a_1}} {\rm{erf}}\left( \frac{1+2a_1  {\rm{ln}}\left(a_2(\lambda_m-1) \right)}{2\sqrt{\pi}}\right) + C.
	\label{eq:lcs21}
\end{equation}
The integration constant $C$ is determined with the help of the condition $G\left(\lambda_m=1\right)=0$. The argument of the error function takes the logarithmic value $-\infty$ for $\lambda_m=1$. The error function then takes the value $-1$ in this argument. As a result, the integration constant follows
\begin{equation}
	C = \frac{a_0 a_2 \rm{exp}\left(\frac{1}{4a_1} \right)\sqrt{\pi}}{2\sqrt{a_1}},
	\label{eq:lcs22}
\end{equation}
after inserting  integration constant in equation \eqref{eq:lcs21} results in cumulative distribution function $G(\lambda_m)$
\begin{equation}
	G(\lambda_m) = \frac{a_0 a_2 \rm{exp}\left(\frac{1}{4a_1} \right)\sqrt{\pi}}{2\sqrt{a_1}}\left(1+ \rm{erf}\left( \frac{1+2a_1  \rm{ln}\left(a_2(\lambda_m-1) \right)}{2\sqrt{\pi}}\right)\right).
	\label{eq:lcs23}
\end{equation}
\section{Micromechanical network viscoelastic material model}
The micromechanical material laws defined so far are suitable for investigating mechanical behaviour. The viscoelastic behaviour is modelled using the rheological theory based on the continuum mechanics.
\begin{figure}[H]
	\centering
	\vspace{4mm}
	\def\svgwidth{0.8\textwidth}
	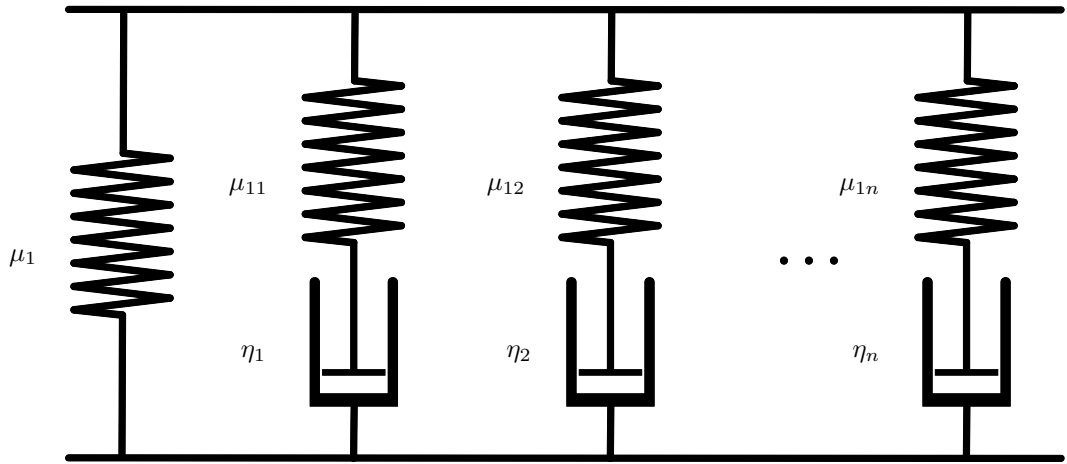
	\vspace*{2mm}	
	\caption{Rheological model of the viscoelasticity with $n$ Maxwell elements.}
	\label{finviskoelast40}
\end{figure}
The rheological theory to describe the viscoelastic behaviour considers the spring element connected in parallel to $j=1,2,...,n$ Maxwell elements (see Fig. \ref{finviskoelast40}). The spring element reflects the basic elasticity, and the Maxwell element represents inelastic material behaviour due to viscosity. The Maxwell element is formed by connecting the spring and dashpot in series. In a physical sense, Maxwell elements approximate the continuous relaxation spectrum of the viscoelastic material. Hence, the viscoelastic behaviour can be represented by a sufficient number of Maxwell elements. The free energy of the material is additively decomposed into volumetric and isochoric parts due to an assumption of a nearly incompressible viscoelastic behaviour. The isochoric free energy is calculated as the sum of equilibrium and non-equilibrium parts.
\begin{equation}
	W = W_{\rm vol} \left(J,\lambda_m \right) + W_{\rm eq} \left(\bar{\rm I}_1^{\bar{\mathbf{B}}},\lambda_m \right) + \sum_{j=1}^{n} W_{\rm neq}\left(\bar{\rm I}_1^{\bar{\mathbf{B}}^j_e},\lambda_m\right),
	\label{eq:totener}
\end{equation}
where $W_{\rm vol}$ is the volumetric part. $W_{\rm eq}$ and $W_{\rm neq}$ are the equilibrium and non-equilibrium parts of the free energy. 
\subsection{Kinematics of finite-strain deformation}
In the framework of the finite strain deformation of a nearly incompressible behaviour, the deformation gradient $\mathbf{F}$ is multiplicatively decomposed into volumetric $\mathbf{F}_{\rm vol}$ and isochoric $\mathbf{F}_{\rm iso}$ parts
\begin{equation}
	\mathbf{F}=\mathbf{F}_{\rm vol}\cdot\mathbf{F}_{\rm iso},\hspace{2mm}{\rm with} \hspace{2mm} \mathbf{F}_{\rm vol} = J^{1/3}\mathbf{I},\hspace{2mm} {\rm and} \hspace{2mm} \mathbf{F}_{\rm iso} = J^{-1/3} \mathbf{F}.
	\label{defogr}
\end{equation}
The isochoric deformation tensor is further multiplicatively decomposed into elastic $\mathbf{F}_{\rm iso}^e$ and inelastic $\mathbf{F}_{\rm iso}^i$ parts
\begin{equation}
	\mathbf{F}^e_{\rm iso} = \left(\det\mathbf{F}_e\right)^{1/3}\mathbf{F}_e, \hspace{3mm} {\rm and} \hspace{3mm}\mathbf{F}^i_{\rm iso} = \left(\det\mathbf{F}_i\right)^{1/3}\mathbf{F}_i.
\end{equation}
The left Cauchy-Green deformation tensor of the elastic and inelastic parts are needed to compute the constitutive equations and are introduced as 
\begin{equation}
	\bar{\mathbf{B}} = J^{-2/3}\mathbf{B},\hspace{3mm}{\rm and} \hspace{3mm} \bar{\mathbf{B}}_{e}^j \,=  \bar{\mathbf{F}}^j \cdot( \bar{\mathbf{C}}_{i}^j)^{-1} \cdot\left(\bar{\mathbf{F}}^j\right)^T,
\end{equation}
where $\bar{\mathbf{B}}_{e}^j$ and $\bar{\mathbf{C}}_{i}^j$ are the elastic and inelastic deformations of the Maxwell element. The inelastic deformation $\bar{\mathbf{C}}_{i}^j$ is evaluated with the rate-dependent evolution equation \cite{Haupt,Seldan}  introduced as 
\begin{equation}
	\begin{aligned}
			\dot{\bar{\mathbf{C}}}_i^j\,&=\,\frac{4}{\eta_j} \frac{\partial W_{\rm neq}^j} {\partial{{\rm{I}}_{\bar{\mathbf{B}}_e^j}}} \left[\bar{\mathbf{C}} - \frac{1}{3}{\rm{tr}}\left(\bar{\mathbf{C}}\cdot\left(\bar{\mathbf{C}}_i^j\right)^{-1} \right)\bar{\mathbf{C}}_i^j\right].
	\end{aligned}
\end{equation}
The right Cauchy-Green deformation of $j^{th}$ Maxwell element required in the evaluation of evolution equation for $\bar{\mathbf{C}}_{i}^j$ are introduced as 
\begin{equation}
	\bar{\mathbf{C}} = J^{-2/3}\mathbf{C},\hspace{3mm}{\rm and}\hspace{3mm}\bar{\mathbf{C}}_i^j = \left(\bar{\mathbf{F}}_i^j\right)^T\cdot\bar{\mathbf{F}}_i^j.
\end{equation}
\section{Variational approach for phase-field fracture}
\subsection{Regularization of phase-field formulation}
The phase-field formulation is based on the diffusive crack equilibrium defined as the crack integral. The idea behind the phase-field fracture model is to minimize the free energy obeying the kinetically admissible displacement field. The free energy of the coupled formulation is introduced as
\begin{equation}
		W\left(\bar{{\rm I}}_1^{\bar{\mathbf{B}}},\bar{{\rm I}}_1^{\bar{\mathbf{B}}_e^j}, J, \lambda_c, \phi\right) = W_b\left(\bar{{\rm I}}_1^{\bar{\mathbf{B}}},\bar{{\rm I}}_1^{\bar{\mathbf{B}}_e^j}, J, \lambda_c, \phi\right) + W_s\left(\phi\right).
		\label{eq:toten}
\end{equation}
where $W_b$, $W_s$ refers to the stored mechanical energy and the surface energy associated with the formation of a crack. Crack and displacement field are studied in time $T\subset \mathbb{R}$ for a three-dimensional spatial domain $\rm \Omega \subset \mathbb{R}^\delta$ in the material configuration with dimension $\delta \in \left\{1,2,3\right\}$. The time-dependent phase-field variable is denoted as
	\begin{equation}
		\phi =
		\begin{cases}
			{\rm\Omega}\times T \rightarrow \left[0,\,1\right]\\
			\left({\rm x}, t\right)\rightarrow \phi\left({\rm x}, t\right)
		\end{cases}, 
		\label{eq:phi}      
	\end{equation}
	and the displacement field $\mathbf{u}$ is defined as
	\begin{equation}
		\mathbf{u} =
		\begin{cases}
			{\rm\Omega}\times T \rightarrow \left[0,\,1\right]\\
			\left({\rm x}, t\right)\rightarrow \mathbf{u}\left({\rm x}, t\right)
		\end{cases}.
		\label{eq:disp}      
	\end{equation}
	The stored mechanical energy is considered to evaluate the degradation in mechanical energy due to the crack propagation and is introduced as 
	\begin{equation}
		W_b\left(\bar{{\rm I}}_1^{\bar{\mathbf{B}}},\bar{{\rm I}}_1^{\bar{\mathbf{B}}_e^j}, J, \lambda_c, \phi\right) = g(\phi)W_0\left(\bar{{\rm I}}_1^{\bar{\mathbf{B}}},\bar{{\rm I}}_1^{\bar{\mathbf{B}}_e^j}, J, \lambda_c\right),
		\label{eq:bulk}
	\end{equation}
	where $W_0$ and $g(\phi)$ are the viscoelastic free energy defined and the degradation function. Degradation of the bulk energy needs to satisfy necessary conditions with the phase-field variable
	\begin{equation}
		\begin{aligned}
			&W_b\left(J,\bar{{\rm I}}_1^{\bar{\mathbf{B}}},\bar{{\rm I}}_1^{\bar{\mathbf{B}}_e^j},  \lambda_c,\phi= 1\right) = W_0\left(J,\bar{{\rm I}}_1^{\bar{\mathbf{B}}},\bar{{\rm I}}_1^{\bar{\mathbf{B}}_e^j}, \lambda_c\right),\\ 
			&W_b\left(J,\bar{{\rm I}}_1^{\bar{\mathbf{B}}},\bar{{\rm I}}_1^{\bar{\mathbf{B}}_e^j},  \lambda_c, \phi=0\right) = 0, \hspace{1mm} \partial W_b\left(J, \bar{{\rm I}}_1^{\bar{\mathbf{B}}},\bar{{\rm I}}_1^{\bar{\mathbf{B}}_e^j},  \lambda_c, \phi\right) < 1,\\
			&{\rm and}\,\,\, \partial W_b\left(J,\bar{{\rm I}}_1^{\bar{\mathbf{B}}},\bar{{\rm I}}_1^{\bar{\mathbf{B}}_e^j},  \lambda_c, \phi=0\right) = 0.
		\end{aligned}	
	\end{equation}  
	The surface energy density is based on a well-known energy balance of Griffith. According to this, the energy release per unit progress of an existing crack must be equal to the energy consumption in creating a new surface, and this quantity is called fracture energy $W_s$
	\begin{equation}
		W_s(\phi) = \int\limits_{\rm\Gamma} E_c \rm d \Gamma.
	\end{equation}
	where $E_c$ is the critical energy release rate. The crack surface $\partial\Gamma$ is an unknown entity and the solution is not possible. Therefore the crack surface $\partial\Gamma$ is regulated by means of a function $\gamma$ 
	\begin{equation}
		W_s(\phi) = \int_{\rm\Omega} E_c \gamma(\phi, \nabla \phi).
		\label{eq:phasefun}
	\end{equation}
	For a defined body $\rm \Omega$, the crack area density function $\gamma$ is regularized using an elliptic function \cite{Ambrosio,Ambrosio92} 
	\begin{equation}
		\gamma\left(\phi, \, \nabla\phi\right)=\frac{1}{2\ell_f} \left(1-\phi\right)^2 + \frac{\ell_f}{2}|\nabla\phi|^2.
		\label{eq:regula}
	\end{equation}
	After substituting equations \eqref{eq:bulk}, \eqref{eq:phasefun} and \eqref{eq:regula} in equation \eqref{eq:toten} results in the total potential energy  \cite{Bourdin2008,Francfort}
	\begin{equation}
		\begin{aligned}
			W\left(J, \bar{{\rm I}}_1^{\bar{\mathbf{B}}},\bar{{\rm I}}_1^{\bar{\mathbf{B}}_e^j}, \lambda_c, \phi\right)= \int\limits_{\rm \Omega} g(\phi) W_0\left(J,\bar{{\rm I}}_1^{\bar{\mathbf{B}}},\bar{{\rm I}}_1^{\bar{\mathbf{B}}_e^j},  \lambda_c\right){\rm dV}
			+ \int\limits_{\rm \Omega} E_c \left[\frac{1}{2\ell_f} \left(1-\phi\right)^2 + \frac{\ell_f}{2}|\nabla\phi|^2\right]{\rm dV}.
			\label{eq:totene}
		\end{aligned}
	\end{equation} 
	\subsection{Degradation function}
	The energetic degradation function $g(\phi)$ is important as it couples the crack phase-field and the mechanical fields. More specifically, it determines how the stored energy function responds to changes in the crack phase-field. The energy degradation function must satisfy the following conditions:
	\begin{itemize}
		\item $g(0) = 0$ damaged material and $g(1) = 1$ undamaged material;\\
		\item $g'(\phi)=\frac{\partial g}{\partial \phi}<1$ need to be a monotonically decay function;\\
		\item $g'(0)=0$ controls the contribution of bulk energy in the evolution of phase-field crack\\
	\end{itemize}
	The third property arises from the crack driving force term $g'(\phi)$ in the evolution law to guarantee the localisation band does not exhibit orthogonal increment. A second-order degradation function is considered along with a regularisation variable \cite{Bourdin2000} to guarantee convergence.
	\begin{equation}
		g(\phi) = (1-\zeta)\phi^{2}+\zeta
		\label{eq:degfun}
	\end{equation}
	The regularisation parameter $\zeta$ must be small to avoid overestimating the mechanical energy and underestimation of the crack energy \cite{BORDEN201277,Braides2000,Bourdin2000}.
	\subsection{Micromechanically motivated fracture energy}
	The early work carried out to formulate the fracture energy of a network polymer is based on the breakage of the individual chains formed of monomers of equal length. The fracture energy $g_{\rm chain}$ of a single chain is introduced as a multiplication between the energy $U$ required to break a monomer with the number of monomer units $N$ needed to form a single chain.
	\begin{equation}
		g_{\rm chain} = NU.
	\end{equation}
	The macroscopic definition of fracture energy of the diffusive crack in a polymer network considers the number of chains $n_{\Gamma}$ tend to break in a unit cross-sectional area. The number of chains $n_{\Gamma}$ tend to break in a unit cross-sectional area is formulated using number of chains per unit volume $n$ for mean end-to-end distance of chain $r_0$ and shorter and longer chain distribution is considered by using cumulative distribution function $G(\lambda_{m})$
	\begin{equation}
		n_{\Gamma} = \frac{1}{2}\left(r_0 G(\lambda_{m})\right)n.
		\label{eq:ngama}
	\end{equation}
	The macroscopic fracture energy $E_c$ of the crosslinked polymer is introduced by multiplying the fracture energy $g_{\rm chain}$ of a uniform chain with the number of chains $n_{\Gamma}$ that break in a unit cross-sectional area 
	\begin{equation}
		E_c = n_{\Gamma}g_{\rm{chain}}.
		\label{eq:ec}
	\end{equation}
	For an assumed length of $N$ monomer units of equal length $l$, the mean end-to-end distance of each chain is computed as $r_0=\left(8N/3\pi\right)^{\frac{1}{2}}l$. Substituting $r_0$ in equation \eqref{eq:ngama} results in $n_{\Gamma} = G(\lambda_{m})\left(2N/3\pi\right)^{\frac{1}{2}}ln$. Furthermore, substituting $n_{\Gamma}$ and $g_{\rm chain}$ in equation \eqref{eq:ec} leads to the macroscopic fracture energy
	\begin{equation}
		E_c = \left(G(\lambda_{m})\sqrt{\frac{2}{3\pi}}\right)\,lUnN^\frac{3}{2} = G(\lambda_{m}) g_c.
		\label{effener}
	\end{equation}
	\subsection{Variational approach in solving phase-field damage}
	The variational formulation of the free energy function is derived for the field variables $\left(\mathbf{u},\phi\right)$ by introducing the test functions of phase-field $\delta\phi$ and displacement $\delta\mathbf{u}$ fields
	\begin{equation}
		\partial W = \left(\frac{\partial W}{\partial \mathbf{u}}\right)\colon \delta \mathbf{u} + \left(\frac{\partial W}{\partial \phi}\right)\colon\delta\phi.
	\end{equation}
	For the free energy function introduced in equation \eqref{eq:totene} that is based on the finite-strain phase-field damage theory follows 
	\begin{equation}
		\begin{aligned}
			\partial W = \int\limits_{\rm \Omega} \left\{g(\phi) \mathbf{T} \colon{\rm grad ^s}\delta\mathbf{u} + g'(\phi)\, \delta\phi W_0 \right\}{\rm dV} + 
			 \int\limits_{\rm \Omega} \left\{E_c \left[-\frac{1}{\ell_f} \left(1-\phi\right) \delta\phi + \ell_f \nabla \phi  \nabla \delta\phi\right]\right\}{\rm dV},
			\label{eq:variational}
		\end{aligned}
	\end{equation}
	where ${\rm grad ^s}\delta\mathbf{u} = \frac{1}{2}\left[{\rm grad}\delta\mathbf{u} + \left({\rm grad}\delta\mathbf{u}\right)^T\right]$ is required for symmetry in the definition of the stress tensor. After substituting the degradation function introduced in equation \eqref{eq:degfun} and its derivative $g'(\phi)$ in equation \eqref{eq:variational} leads to
	\begin{equation}
		\begin{aligned}
			\partial W = \int\limits_{\rm \Omega}& \left\{\left[(1-\zeta)\phi^{2}+\zeta\right] \mathbf{T}\colon{\rm grad ^s}\delta\mathbf{u}\right\}{\rm dV} + \\
			\int\limits_{\rm \Omega}& \left\{2\left(1-\zeta\right) \phi\, \delta\phi W + E_c \left[-\frac{1}{\ell_f} \left(1-\phi\right) \delta\phi + \ell_f \nabla \phi  \nabla \delta\phi\right]\right\}{\rm dV}.
		\end{aligned}\label{cons}
	\end{equation}
	The strong forms of the governing equations of the phase-field damage model are derived by applying the Gauss theorem over the equation \eqref{cons}. The governing equations are represented as
	\begin{subequations}
		\begin{align}
			{\rm Div}\left(\left[(1-\zeta)\phi^{2}+\zeta\right] \mathbf{T}\right) &= \mathbf{0} \label{eq:mech}\\
			\underbrace{2\left(1-\zeta\right) \phi\, W}_{\rm driving\, force} + \underbrace{E_c \left[-\frac{1}{\ell_f} \left(1-\phi\right) + \ell_f \Delta \phi \right]}_{\rm resistance\, to\, crack} &= 0.
			\label{eq:phase}
		\end{align}
	\end{subequations}
	Equation \eqref{eq:mech} corresponds to the balance of momentum to evaluate the viscoelastic response, and equation \eqref{eq:phase} is the phase-field evolution of the diffusive crack propagation. The first term of the phase-field evolution is responsible for the energetic driving crack in the continuum domain and the second term refers to the geometric resistance to the propagation of the crack. The time derivative of the phase-field variable needs to follow $\dot{\phi}\le 0$ to avoid crack irreversibility.
	\subsection{Boundary conditions}
	The governing equations to determine the fracture phase-field $\phi$ and the displacement field $\mathbf{u}$ of the body are introduced in the equations \eqref{eq:mech}-\eqref{eq:phase} need to satisfy essential and necessary boundary conditions.  To this end, the boundaries $\partial \rm \Omega$ are decomposed into displacement and damage fields and have to satisfy the conditions
	\begin{equation}
		\begin{aligned}
			\partial{\rm \Omega}=\partial{\rm \Omega}_{\mathbf{u}}^D\cup\partial{\rm \Omega}_{\mathbf{t}}^N &\,\, {\rm and}\,\, \partial{\rm \Omega}=\partial{\rm \Omega}_{\phi}^D\cup\partial{\rm \Omega}_{\nabla\phi}^N\\
			\partial{\rm \Omega}_{\mathbf{u}}^D\cap\partial{\rm \Omega}_{\mathbf{u}}^N = \varnothing &\,\, \text{and} \,\, \partial{\rm \Omega}_{\phi}^D\cap\partial{\rm \Omega}_{\nabla\phi}^N = \varnothing,
		\end{aligned}
	\end{equation}
	where $\partial{\rm \Omega}_{\mathbf{u}}^D$ and $\partial{\rm \Omega}_{\mathbf{t}}^N$ are the Dirichlet and Neumann boundaries of the displacement field. $\partial{\rm \Omega}_{\phi}^D$ and $\partial{\rm \Omega}_{\nabla\phi}^N$ are the Dirichlet and Neumann boundaries of the damage field. The displacement $\mathbf{u}$ and traction $\mathbf{t}$ on the boundaries are the Dirichlet and the Neumann boundary conditions  
	\begin{equation}
		\mathbf{u}\left(\mathbf{x},t\right) = \mathbf{u}_D\left(\mathbf{x},t\right) \,\, \text{at} \hspace{2mm} \mathbf{x} \in \partial{\rm \Omega}_{\mathbf{u}}^D \,\, {\rm and} \,\, \mathbf{T}\cdot\mathbf{n} = \mathbf{t} \,\, {\rm on} \,\,\partial{\rm \Omega}_{\mathbf{t}}^N.
	\end{equation}
	For the phase-field damage, the cracked region is constrained by the Dirichlet and the Neumann boundary conditions on the crack surface with
	\begin{equation}
		\phi\left(\mathbf{x},t\right) = 0 \hspace{2mm} \text{at} \hspace{2mm} \mathbf{x} \in \partial{\rm \Omega}_{\phi}^D\,\,{\rm and}\,\,\nabla\phi\cdot\mathbf{n}=0\,\,{\rm on}\,\,\partial{\rm \Omega}_{\nabla\phi}^N.
	\end{equation}
	\section{Numerical implementation}
	Analytical solution of the strong forms are often impossible. Therefore, the finite element method is chosen as an approximate method to solve the strong form by deriving weak forms. The numerical solution of the coupled formulation is implemented in an open-source finite element library deal.II \cite{DanielA,Bangerth}.
	\subsection{Finite element implementation of variational principle}
	To develop a numerical scheme of a coupled system of equations using the finite element method, it is convenient to express the residuals of the partial differential equations in the weak forms. The weak forms of equations \eqref{eq:mech}-\eqref{eq:phase} are introduced as
	\begin{equation} \label{eq:15}
		\begin{aligned}
			\mathbf{r}^{\mathbf{u}}_{i} = \int\limits_{\rm \Omega} \left\{\left[(1-\zeta)\phi^{2}+\zeta\right] \mathbf{T} \colon{\rm grad ^s}\delta\mathbf{u}\right\}{\rm dV} &= \mathbf{0}\\
			{\rm r}^{\phi}_{i} = \int\limits_{\rm \Omega} \left\{2\left(1-\zeta\right) \phi\, \delta\phi W + E_c \left[-\frac{1}{\ell_f} \left(1-\phi\right) \delta\phi + \ell_f \nabla \phi  \nabla \delta\phi\right]\right\}{\rm dV} &= 0.
		\end{aligned}
	\end{equation}
	In the context of the finite element method, the primary variables are displacement $\mathbf{u}$ and phase-field variable $\phi$ approximated as 
	\begin{equation}
		\mathbf{u} = \sum_{i=1}^{{\rm n}_{\rm ele}} \mathbf{N}_i^{\mathbf{u}} \mathbf{u}_i, \hspace{1 cm} \phi = \sum_{i=1}^{{\rm n}_{\rm ele}} {N}_i {\phi}_i,
	\end{equation}
	where $\mathbf{N}_i^{\mathbf{u}}$ is given by:
	\begin{equation} \label{eq:18}
		\mathbf{N}_i^{\mathbf{u}} = \begin{bmatrix} 
			{\rm N}_i & 0 & 0 \\ 
			0 & {\rm N}_i & 0 \\ 
			0 & 0 & {\rm N}_i\\ 
		\end{bmatrix}.
	\end{equation}
	In the Equation \eqref{eq:18}, $N_i$ denotes the shape functions of the elements at the quadrature points associated with the respective nodes. $\mathbf{u}_i = \left({\rm u}_x,{\rm u}_y, {\rm u}_z\right)^T$ is the displacement field and $\phi_i$ is the phase-field damage variable. Consequently, the derivatives of the field variables are introduced as
	\begin{equation} \label{eq:19}
		{\rm grad}\, \mathbf{u} = \sum_{i=1}^{{\rm n}_{\rm ele}} \mathbf{B}_i^{\mathbf{u}} {\mathbf{u}}_i\,\,\,\,\,\,{\rm and} \,\,\,\,\,\,{\rm grad}\, \phi = \sum_{i=1}^{{\rm n}_{\rm ele}} {B}_i^{\phi} {\phi}_i.
	\end{equation}
	Herein, we introduced the $\mathbf{B}$ matrix at the nodes given by
	\begin{equation} \label{eq:20}
		\mathbf{B}_i^{\mathbf{u}} = \begin{bmatrix} 
			{\rm N}_{i,x} & 0 & 0\\ 
			0 & {\rm N}_{i,y} & 0\\ 
			0 & 0 & {\rm N}_{i,z}\\
			{\rm N}_{i,y} & {\rm N}_{i,x} & 0\\
			0 & {\rm N}_{i,z} & {\rm N}_{i,y} \\
			{\rm N}_{i,z} & 0 & {\rm N}_{i,x}\\
		\end{bmatrix}, \hspace{1 cm}
		\mathbf{B}_i^{\phi} = \begin{bmatrix} 
			{\rm N}_{i,x}\\ 
			{\rm N}_{i,y} \\ 
			{\rm N}_{i,z} \\
		\end{bmatrix},
	\end{equation}
	where, $N_{i,x}$, $N_{i,y}$ and $N_{i,z}$ are the derivatives of the shape functions. Similarly, approximation of the virtual displacements are computed as
	\begin{equation} \label{eq:21}
		\begin{split}
			\delta \mathbf{u} = \sum_{i=1}^{\text{n}_\text{ele}} \mathbf{N}_i^{\mathbf{u}} \delta \mathbf{u}_i \hspace{1.8 cm} \delta \phi = \sum_{i=1}^{\text{n}_\text{ele}} {N}_i \delta {\phi}_i \\
			{\rm grad^s}\delta\mathbf{u} = \sum_{i=1}^{\text{n}_\text{ele}} 
			\mathbf{B}_i^{\delta \mathbf{u}} \mathbf{u}_i \hspace{1. cm} {\rm grad}\, \delta \phi = \sum_{i=1}^{\text{n}_\text{ele}} {B}_i^{\phi} {\delta \phi}_i.
		\end{split}
	\end{equation}
	The weak forms of the residuals given in equation \eqref{eq:15} are introduced with the above-mentioned finite element expressions for the field variables as
	\begin{equation}
		\begin{aligned}
			\mathbf{r}^{\mathbf{u}}_{i} = \int\limits_{\rm \Omega} \left\{\left[(1-\zeta)\phi^{2}+\zeta\right] \mathbf{T} \colon(\mathbf{B}_i^{\mathbf{u}})^\text{T}\right\}{\rm dV} &= \mathbf{0}.\\
			\hspace{-1mm}{\rm r}^{\phi}_{i} =\!\! \int\limits_{\rm \Omega}\! \left\{2\left(1-\zeta\right) \phi\, N_i W \!+ E_c \left[-\frac{1}{\ell_f} \left(1-\phi\right) N_i \!+\! \ell_f ( \mathbf{B}_i^{\phi})^{T}  \nabla \delta\phi\right]\!\right\}\!{\rm dV} &= 0.
		\end{aligned}
	\end{equation}
	The weak form of the coupled system of equations yields the static equilibrium in the form of a non-linear differential vector equation due to geometrical non-linearity because of the large deformations. The linearised approximation of the non-linear governing equations is solved using Newton's method with an increment scheme given as:
	\begin{equation}
		{\begin{bmatrix} 
				\mathbf{{K}}^\mathbf{uu} & \mathbf{{K}}^{\mathbf{u}\phi} \\ 
				\mathbf{{K}}^{\phi\mathbf{u}} & \mathbf{{K}}^{\phi \phi} \\ 
		\end{bmatrix}} 
		{\begin{bmatrix} d \mathbf{\mathbf{u}}\\ 
				d {{{\phi}}} \\ 
		\end{bmatrix}} = {\begin{bmatrix} -\mathbf{{R}}_{\mathbf{u}} (\mathbf{u}_{\textrm{i}}) \\ 
				-\mathbf{{R}}_{\phi} ({\phi}_{\textrm{i}})
		\end{bmatrix}}.
	\end{equation}
	The directional derivatives of the spatial tangent tensor are calculated to solve the finite element problem as:
	\begin{equation}
		\begin{aligned}
			&\mathbf{K}_{i,j}^{\mathbf{uu}} = \frac{\partial \mathbf{r}_i^\mathbf{u}}{\partial \mathbf{u}_j} = \int\limits_{\rm \Omega}  \left( \left( 1 - \zeta \right) \phi^2 + \zeta \right)\mathbf{B}^{\mathbf{u}}_i \colon\boldsymbol{\kappa}\colon \mathbf{B}^{\mathbf{u}}_j {\rm dV}, \\
			&\mathbf{K}_{i,j}^{\mathbf{u}\phi} = \frac{\partial \mathbf{r}_i^\mathbf{u}}{\partial \phi_j} = \int\limits_{\rm \Omega} 2(1-\zeta)\phi \mathbf{B}^{\mathbf{u}}_i\colon \mathbf{T}^T N_j {\rm dV},\\
			&\mathbf{K}_{i,j}^{\phi\mathbf{u}} = \frac{\partial \mathbf{r}_i^{\phi}}{\partial \mathbf{u}_j} = \int\limits_{\rm \Omega} 2(1-\zeta)\phi N_i \mathbf{T}^T \colon\mathbf{B}^{\mathbf{u}}_j {\rm dV},\\
			&\mathbf{K}_{i,j}^{\phi \phi} = \frac{\partial \mathbf{r}_i^{\phi}}{\partial \phi_j} = \int\limits_{\rm \Omega}\!\! \left\lbrace \! \left(1 - \zeta \right) W N_i N_j \!+ \!E_c \left[\frac{1}{\ell_f} N_i N_j + \ell_f (\mathbf{B}^{\phi}_i)^{\text{T}}\colon \mathbf{B}^{\phi}_j \right]\!\right\rbrace {\rm dV}. \\
		\end{aligned}
	\end{equation}
	The tangent tensor $\boldsymbol{\kappa}$ of the network model is evaluated as the sum of tangent tensors of volumetric $\boldsymbol{\kappa}_{\rm vol}\left(J\right)$, equilibrium $\boldsymbol{\kappa}_{\rm eq}\left(\bar{\mathbf{B}}\right)$ and non-equilibrium parts $\boldsymbol{\kappa}_{\rm neq}\left(\bar{\mathbf{B}}_e^j\right)$ of $j=1,2,...,n$ Maxwell elements.
	\section{Investigation of phase-field fracture}
\begin{figure}[H]
		\centering
		\def\svgwidth{0.8\textwidth}
		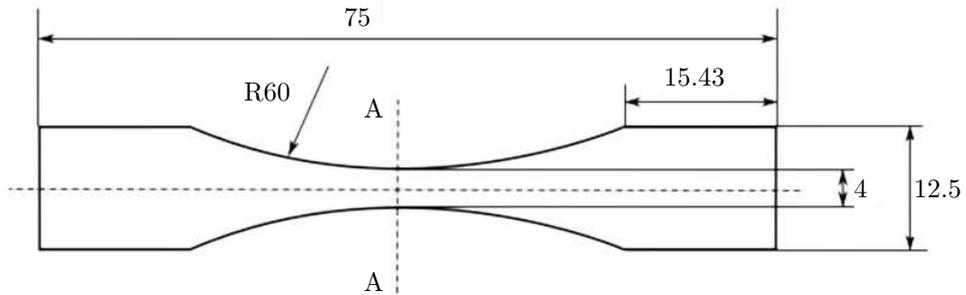
		\caption{Tailored tensile test samples with a necked cross-section at the centre of the sample: all dimensions are in millimetres}
		\label{FIG:2DTail}
	\end{figure}

\begin{figure}[H]
		\centering
		\scalebox{0.9}{\input{tension/BF5e4_EXPVSSIM_0}}
		\caption{Numerical simulation and experimental data comparison of tensile test}
		\label{fig:abs}
	\end{figure}
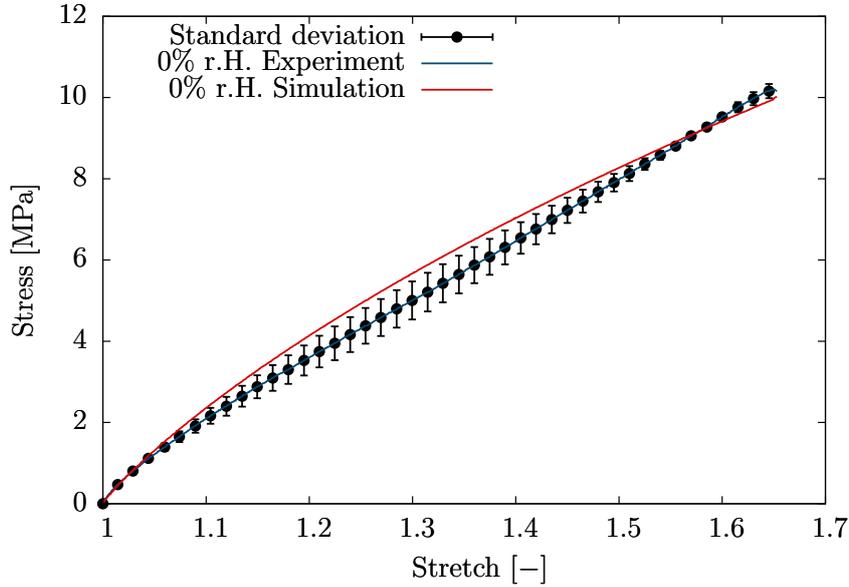

\begin{figure}[H]
		\centering
		\scalebox{.42}{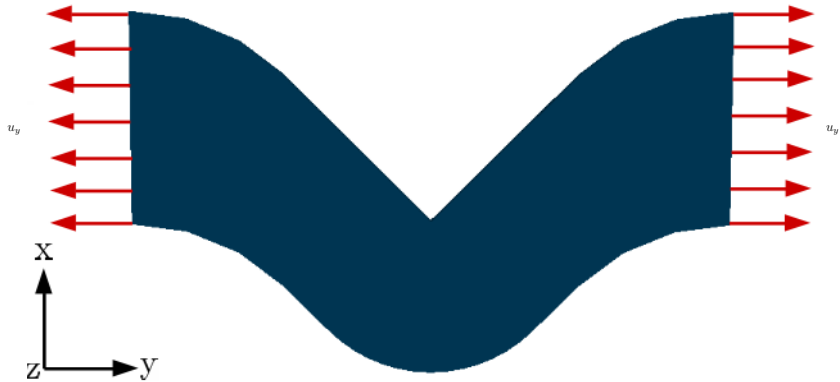}
		\caption{Angular specimen applied with the displacement $u_y$ and $u_x=u_z=0$ as Dirichlet boundary conditions and traction free boundary conditions as tensile loading condition.}
		\label{bcs}
	\end{figure}

\begin{figure}[!htp]
		\centering
		\begin{tabular}{ccc}		
			\subf{\includegraphics[scale=0.7]{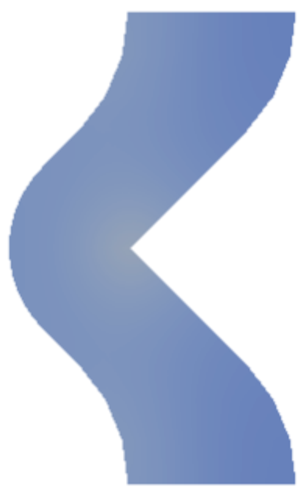}}
			{Time $t= 25\,\rm s$}
			&
			\hspace{10 mm}\subf{\includegraphics[scale=0.7]{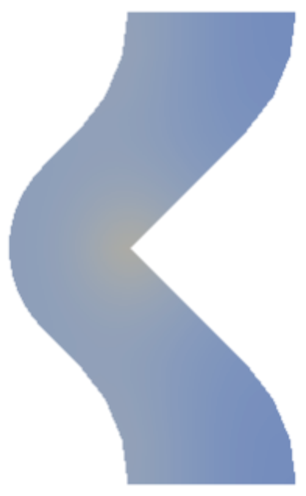}}
			{Time $t= 30\,\rm s$} 
			&
			\multirow{2}{*}{\subf{\def\svgwidth{0.05\textwidth}
					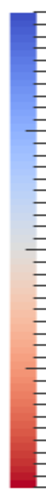
			}}
			\\	
			\\	
			\subf{\includegraphics[scale=0.7]{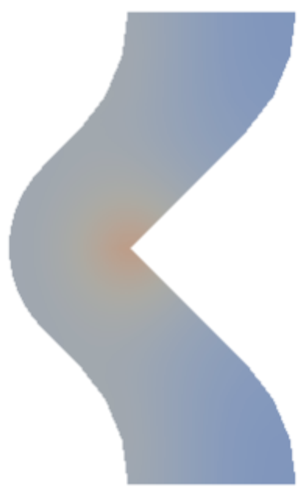}}
			{Time $t= 35\,\rm s$}
			&
			\hspace{10 mm}\subf{\includegraphics[scale=0.7]{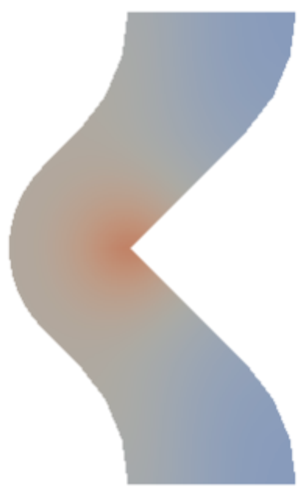}}
			{Time $t= 43\,\rm s$} &
		\end{tabular}
		\caption{Contour plots of the crack propagation at different times of the finite element simulation in the angular sample (DIN ISO 34-1) under tensile loading condition}
		\label{fig:contorplot}
	\end{figure}

\begin{figure}[H]
		\begin{subfigure}{0.5\textwidth}
			\scalebox{.65}{\input{tension/crack2}}
			\caption{Force vs. Time}
		\end{subfigure}%
		\begin{subfigure}{0.5\textwidth}
			\scalebox{.65}{\input{tension/crack2VST}}
			\caption{Force vs. Displacement}\label{FVSDorh}
		\end{subfigure}
		\caption{Tear tests are performed on dry samples of adhesive-A at $60^\circ$C}
		\label{orhdama}
	\end{figure}
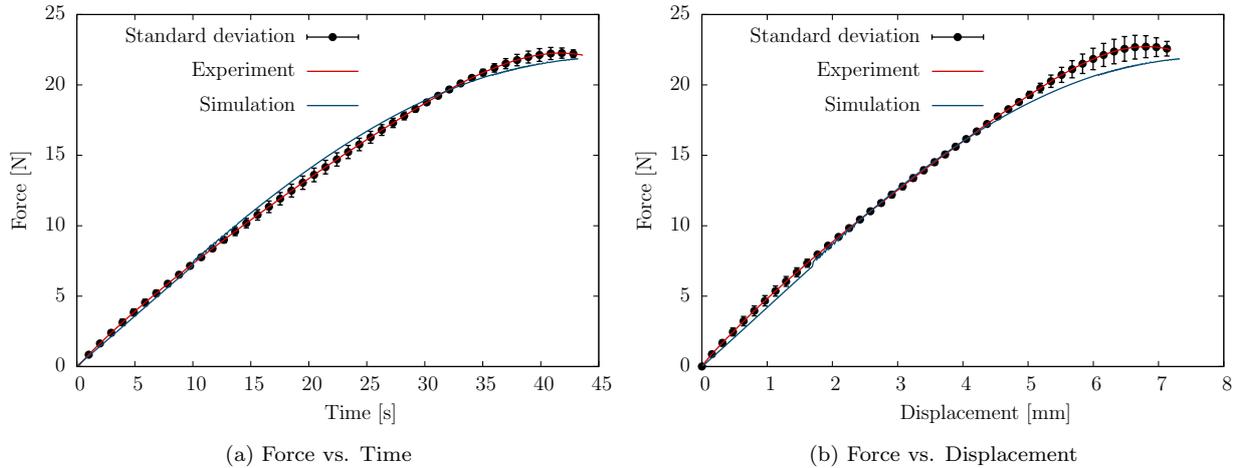

	Micromechanically motivated viscoelastic network model parameters are identified by fitting simulation and uniaxial tensile test curves using a gradient-free algorithm proposed by Nelder \& Mead \cite{Nelder}. Uniaxial tensile tests are performed on the tailored sample shown in Fig \ref{FIG:2DTail} at $60\,^\circ$C. The motivation behind using a tailored sample is to measure deformations locally at a gauge length of $2\,\rm mm$ from the centre of the sample.
	The micromechanically motivated viscoelastic material model is formulated using seven Maxwell element and their parameters are identified with Nelder \& Mead algorithm implemented in MATLAB. 2 mm cross-section from the centre of the tailored sample is considered to apply tensile load boundary conditions to investigate tensile behaviour. The optimal material properties evaluated with the optimization algorithm are summarized in Table \ref{table1}.
	\begin{table}[H]
		\centering
		\caption{Material parameters of micro-mechanical polymer network model at different ambient condition}
		\small\addtolength{\tabcolsep}{10pt}
		\renewcommand{\arraystretch}{1.5}
		\begin{tabular}{|c|c|c|c|} 
			\hline
			\multicolumn{4}{|c|}{Material parameters of micro-mechanical model}                                                                                                                                                                                                                                     \\ \hline
			&                                                                  & \begin{tabular}[c]{@{}c@{}}Relaxation\\ times {[}s{]}\end{tabular} &   \\ \hline
			Equilibrium                                                                    & $c_{10}$ {[}MPa{]}                                               &                                               & 9.183     \\ \hline
			\multirow{7}{*}{Non-equilibrium}                                                & $c_{101}$ {[}MPa{]}                                              & 0.5                                  & 5.223  \\ \cline{2-4} 
			& $c_{102}$ {[}MPa{]}                                              & 10                                                                 & 4.152     \\ \cline{2-4} 
			& $c_{103}$ {[}MPa{]}                                              & 100                                                                & 3.140       \\ \cline{2-4} 
			& $c_{104}$ {[}MPa{]}                                              & 500                                                                & 2.328       \\ \cline{2-4} 
			& $c_{105}$ {[}MPa{]}                                              & 1000                                                               & 1.582      \\ \cline{2-4} 
			& $c_{106}$ {[}MPa{]}                                              & 2500                                                               & 1.131      \\ \cline{2-4} 
			& $c_{107}$ {[}MPa{]}                                              & 5000                                                               & 0.961      \\ \hline
			\multirow{2}{*}{\begin{tabular}[c]{@{}c@{}}Wesslau \\ parameters\end{tabular}} & \begin{tabular}[c]{@{}c@{}}Average chain \\ stretch\end{tabular} &                                                                    & 1.194        \\ \cline{2-4} 
			& \begin{tabular}[c]{@{}c@{}}Polydispersity\\ index\end{tabular}   &                                                                    & 1.001     \\ \hline
		\end{tabular}
		\label{table1}
	\end{table}
	The stress-stretch data from the tension tests are compared with the simulation data obtained with the optimum material parameters to validate the material parameters. Figure \ref{fig:abs} shows the comparison between the experiment and the simulation data with the standard deviation as an error bar. The tension test data plotted in the comparison corresponds to the mean values calculated from the test series consisting of five samples for aged samples at individual humid conditions. The standard deviation in the form of the error bar indicates that the problem is well-posed and sufficient to use the material properties for damage behaviour using the phase-field damage model.
	The proposed finite-strain phase-field damage model to investigate the tear strength of the cross-linked polymers is validated by comparing the simulation and the experimental results. The experimental investigations are performed on the selected polyurethane adhesives based on the DIN ISO 34-1 standards. The test procedure considers an angular specimen under tensile loading conditions to investigate the combination of crack initial and propagation. The geometrical structure of the test specimen is shown in Fig \ref{geom} and the thickness of $2 \, \rm mm$ is considered in the tensile tests. Three batches of dry specimens were manufactured to perform a tear strength test at an isothermal condition of $60\,^\circ$C. The samples were nicked at the angular notch to inflict an initial crack on the sample, and both ends of the samples were clamped (approximately 22 mm in length) to apply uniaxial tensile load until fracture.
	\begin{figure}[H]
		\centering
		\def\svgwidth{0.7\textwidth}
		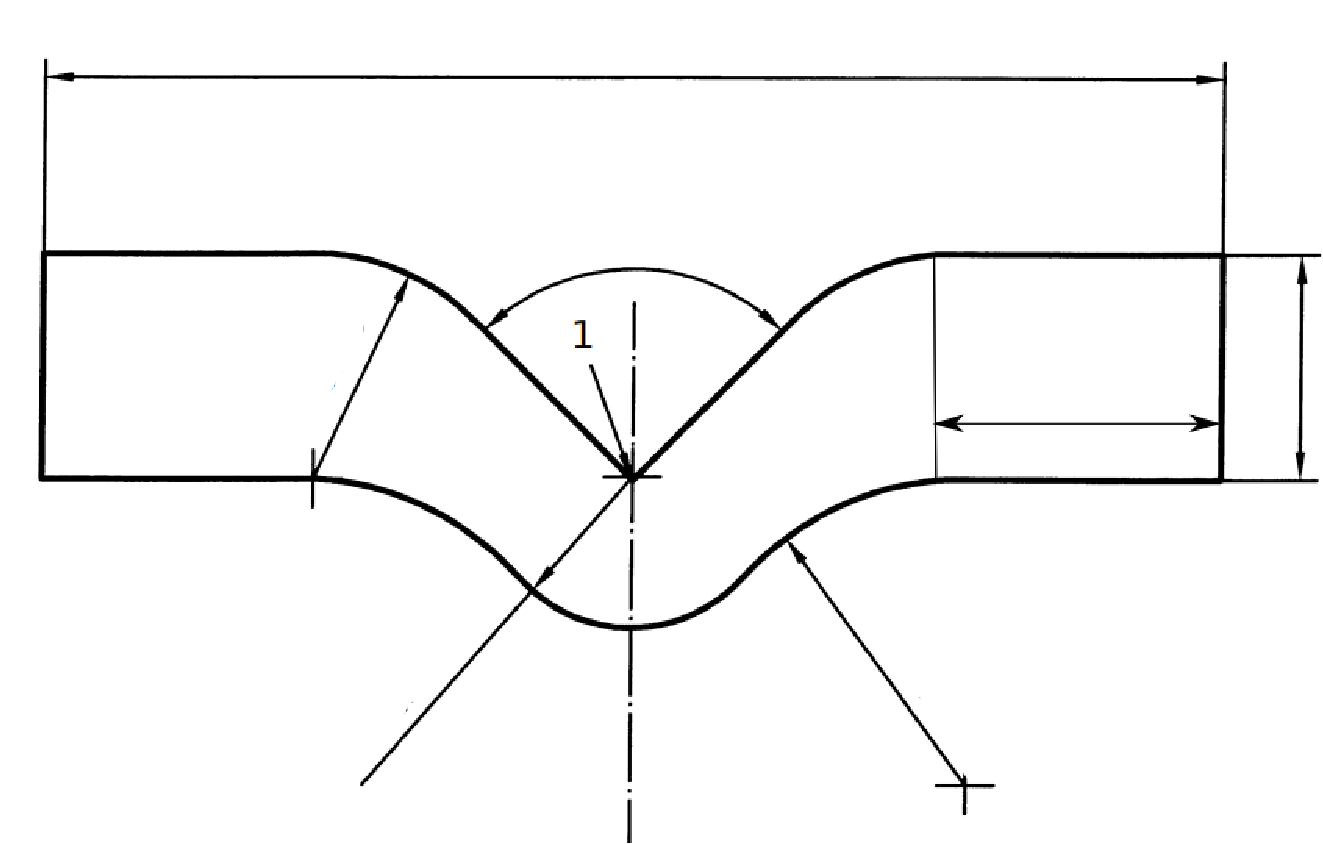
		\caption{Geometry of the angle sample based on DIN ISO 34-1: all dimensions are in millimetres}
		\label{angular}
	\end{figure}
	The geometry of the angular sample is spatially discretised into finite elements for the numerical simulation. The crack initiated initiates and propagates at the angular notch of the sample called as transition zone. The finite element mesh at the transition zone must be sufficiently refined as proposed by Miehe et al. \cite{Miehe2010}. The finite element mesh needs to follow the minimum size of local mesh $h$ to resolve problems involved in crack length $l$ and the local size of the mesh should always be less than the crack length $h<<l$. The relationship of mesh size is needed to resolve $\Gamma$ regularisation on crack. Therefore, The specimen geometry is spatially discretised into finite elements with mesh size at the transition zone as $h=1.23\, \rm mm$ for an initial crack length of $l=9.31\, \rm mm$.
	\begin{figure}[H]
		\centering
		\includegraphics[trim=100 100 100 120,clip,scale=0.55]{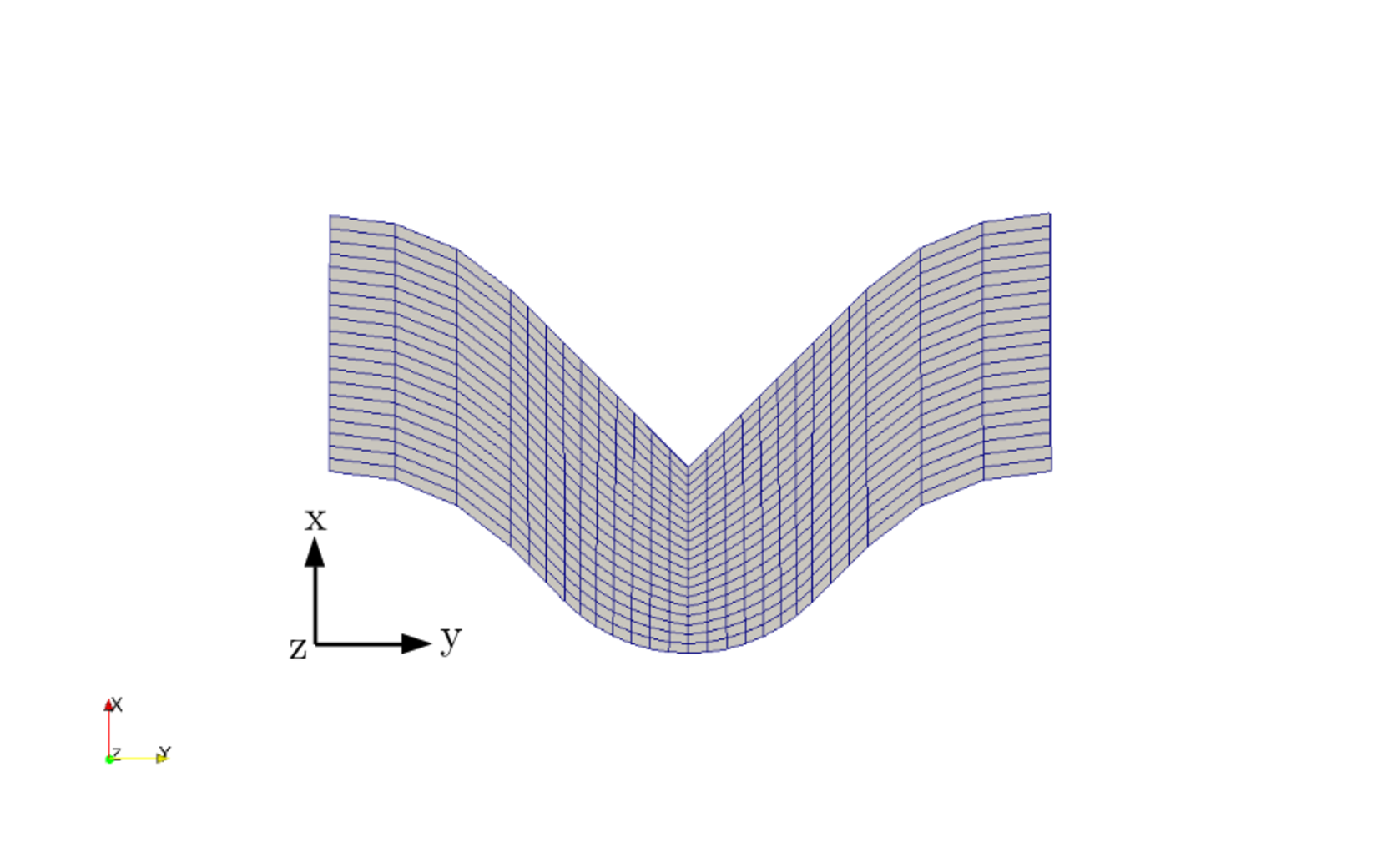}
		\caption{Finite element mesh of angular sample with refined mesh at the transition zone of crack propagation along the centre of the sample.}
		\label{mesh}
	\end{figure}
	The spatially discretized finite element model is defined with the micromechanical viscoelastic parameters listed in Table \ref{table1} to evaluate the viscoelastic behaviour. The critical energy release rate $E_c$ is needed to evaluate the crack propagation using the phase-field evolution equation. The critical energy release rate $E_c$ is evaluated with equation \eqref{effener} by setting $g_c=4.185\,\rm N/mm$. The term $g_c$ corresponds to the energy release rate for the chain length distribution of equal length and is evaluated with energy of monomer as $U = 2.7\rm e^{-13}$, length of chain $l=3.1\,\rm e^{-7}$ and the density of effective chains $n$ for $N$ monomer units in single-chain $nN^{-3/2} = 5\,\rm e^{19}$. The finite element model is solved by defining the Dirichlet boundary conditions for the displacement field at the clamping location and the phase-field variable is set to value $\phi=0$ at the transition zone to define the initial crack. The geometric setup of the three-dimensional boundary value problem is shown in Fig \ref{bcs} with the defined boundary conditions.
	\vspace{9mm}
	The non-linear problem is solved as a monolithic coupled field using the Newton method. The crack evolution to fracture in the deformed configuration is shown in Fig \ref{fig:contorplot}. The crack initiation starts at time $t = 25 \, \rm s$ and propagates until the test piece breaks at $t = 43 \, \rm s$.
	The corresponding force-displacement and force-time curves of the simulation and test results are compared to validate the proposed material model and corresponding properties. The comparison is shown in Fig \ref{orhdama} and the standard deviation is shown with the error bars.
\section{Conclusion}
In the present work, the mechanical behaviour of the crosslinked polymer model is investigated with a micromechanical material model based on the statistic chain mechanics. The proposed material model considers the softening behaviour due to the debonding/breakage of the shorter chains in shorter and longer chain distribution. The proposed micromechanical model is coupled with a phase-field damage model in which the crack energy density is also formulated by considering the shorter and longer chain length distribution. \textbf{Tensile tests} are performed to identify parameters of the micromechanical behaviour, and these mechanical parameters are used in the validation of the proposed micromechanically motivated phase-field damage model. Experiments and finite element simulation are performed on the angular sample to determine \textbf{Tear strength}, and it is established from the comparison of experimental and simulation of tear test that the proposed coupled formulation is robust and efficient in the simulation of crack propagation in crosslinked polymer like polyurethane adhesives. 
\section*{Acknowledgments}
The research project 19730 N "Berechnung des instation\"aren mechanischen Verhaltens von alternden Klebverbindungen unter Einfluss von Wasser auf den Klebstoff" of the research association Forschungsvereinigung Stahlanwendung e.V. (FOSTA), D\"usseldorf was supported by the Federal Ministry of Economic Affairs and Energy through the AiF as part of the program for promoting industrial cooperative research (IGF) on the basis of a decision by the German Bundestag. The experimental investigations are carried out by the project associates from Lehrstuthl f\"ur Angewandte Mechanik, Universit\"at des Saarlandes and Fraunhofer-Institut f\"ur Fertigungstechnik und Angewandte Materialforschung.

\section*{Conflicts of interest}
The authors declare no conflicts of interest.

	\input{PFD_Micromech.bbl}
\newpage
\appendix
\section{Isochoric and volumetric split of the free energy}\label{adddecom}
The free energy density is uniquely decoupled into volume and shape changing components \cite{Holzapfel} for the material which have rubber like behaviour.
\begin{equation}
	W({\bar{\rm I}_1},J,\lambda_m) = W_{{\rm eq}}\left(\bar{{\rm I}}_1^{\bar{\mathbf{B}}},\lambda_m\right) + W_{{\rm neq}}\left(\bar{{\rm I}}_1^{\bar{\mathbf{B}}_e},\lambda_m\right) + W_{{\rm vol}}\left(J,\lambda_m\right),
	\label{micromech}
\end{equation}
where $W_{{\rm eq}}$, $W_{{\rm neq}}$ and $W_{{\rm vol}}$ are the equilibrium, non-equilibrium and volumetric free energies. The equilibrium and non-equilibrium free energy are further interpreted as $W_{{\rm iso}}\left(\bar{{\rm I}}_1,\lambda_m\right)$, where the first invariant $\bar{{\rm I}}_1$ corresponds to either invariants of equilibrium $\bar{{\rm I}}_1^{\bar{\mathbf{B}}}$ or non-equilibrium $\bar{{\rm I}}_1^{\bar{\mathbf{B}}_e}$ parts. To this end, it is necessary to multiplicatively decompose deformation gradient into volumetric $\mathbf{F}_{\rm vol}$ and isochoric parts $\bar{\mathbf{F}}$
\begin{equation}
	\mathbf{F} = \mathbf{F}_{\rm vol}\cdot\bar{\mathbf{F}}\,\,\,{\rm with}\,\,\, \mathbf{F}_{\rm vol} = J^{1/3}\mathbf{I}\,\,\,{\rm resilts\, in}\,\,\, \bar{\mathbf{F}} = J^{-1/3}\mathbf{F},
\end{equation}
and the isochoric left Cauchy-Green tensor is introduced as $\bar{\mathbf{B}}=J^{-2/3} \mathbf{B}$ and the three invariants of the isochoric components are 
\begin{equation}
	\bar{\rm I}_1 = J^{(-2/3)}{\rm I}_1,\,\, \bar{\rm I}_2 = J^{(-4/3)}{\rm I}_2 \,\, \text{and} \,\,\bar{\rm I}_3 = {\rm det}\bar{\mathbf{B}},
\end{equation}
with Jacobian $J = \sqrt{{\rm det}(\mathbf{B})}$. The isochoric form of micromechanical free energy is then formulated with the isochoric first invariant $\bar{\rm I}_1$ and the current stretch $\lambda_c=\sfrac{\sqrt{\bar{{\rm I}}_1}}{\sqrt{3}}$ as
\begin{equation}
	\begin{aligned}
		W_{\rm{iso}}(\bar{\rm{I}}_1,\lambda_m) &= {\rm \Psi}_{\rm{iso}}(\bar{\rm{I}}_1) \left(\int\limits_{1}^{\infty}g_{\rm{iso}}(\lambda_m){\rm{d}} \lambda_m -\int\limits_{1}^{\lambda_c} g_{\rm{iso}}(\lambda_m){\rm{d}} \lambda_m \right)\\
		&= {\rm \Psi}_{\rm{iso}}(\bar{\rm{I}}_1)G_{\rm{iso}}\left(\lambda_c(\bar{\rm{I}}_1) \right)	
		\label{eq:chain-1}
	\end{aligned}
\end{equation}
with the isochoric form of cumulative chain length distribution derived as the function of current chain stretch $\lambda_c(\bar{\rm I}_1) = \sqrt{\bar{\rm I}_1}/\sqrt{3}$ 
\begin{equation}
	G_{\rm{iso}}\left(\lambda_c(\bar{\rm{I}}_1)\right) = 1-\int\limits_{1}^{\lambda_c(\bar{\rm{I}}_1)}g_{\rm{iso}}(\lambda_m)\rm{d} \lambda_m.
	\label{eq:chain-2}
\end{equation}
The distribution function has so far been applied in the formulation of the shape-changing part. However, it is also necessary to consider the distribution function in the volume-changing part since the volumetric part has a huge influence on the tangent stiffness and the stress state. In the case of nearly incompressible materials, the volumetric free energy function is a function of Jacobian since the value of Jacobian is $J=1$. 
\begin{equation}
	\begin{aligned}
		W_{\rm{vol}}(J,\lambda_m) &= {\rm \Psi}_{\rm{vol}}(J) \left( \int \limits_{1}^{\infty} g_{\rm{vol}}(\lambda_m)\rm{d} \lambda_m -\int \limits_{1}^{\lambda_c} g_{\rm{vol}}(\lambda_m) \rm{d} \lambda_m \right)\\
		&= {\rm \Psi}_{\rm{vol}}(J)G_{\rm{vol}}\left(\lambda_c(J) \right)
		\label{eq:chain-10}
	\end{aligned}
\end{equation}
with the volumetric form of cumulative chain length distribution derived as the function of current chain stretch $\lambda_c(J) = J^{1/3}$
\begin{equation}
	G_{\rm{vol}}\left(\lambda_c(J)\right) = 1-\int\limits_{1}^{\lambda_c(J)}g_{\rm{vol}}(\lambda_m)\rm{d} \lambda_m.
	\label{eq:chain-11}
\end{equation}
In order to derive the stress and fourth-order tangent tensors required in the numerical implementation of the material model motivates to derive the first and second-order derivatives of the volumetric and isochoric components of the micromechanical strain energy density. The derivatives of the indefinite integral must be derived with the help of the chain rule due to the implicit dependence of the indefinite integral on the current chain stretch. The first derivative\footnote{First order derivatives of the components: $\frac{\partial(\bullet)}{\partial\lambda_c}=\dot{(\bullet)}$ and $\frac{\partial(\bullet)}{\partial {\bar{\rm I}}_1}=(\bullet)'$\\} follows
\begin{equation}
	G'_{\rm{iso}}\left(\lambda_c(\bar{\rm{I}}_1)\right) = \frac{\partial G_{\rm{iso}}}{\partial\lambda_c}\frac{\partial\lambda_c}{\partial \bar{\rm{I}}_1} =\dot{G}_{\rm{iso}}\lambda_c',
	\label{eq:chain-3}
\end{equation}
and the second derivative\footnote{Second order derivatives of the components: $\frac{\partial^2(\bullet)}{\partial\lambda_c^2}=\ddot{(\bullet)}$ and $\frac{\partial^2(\bullet)}{\partial {\bar{\rm I}}_1^2}=(\bullet)''$} is derived with the product rule as
\begin{equation}
	\begin{aligned}
		G''_{\rm{iso}}\left(\lambda_c(\bar{{\rm{I}}}_1)\right) = \frac{\partial}{\partial \bar{{\rm{I}}}_1}\left(\dot{G}_{\rm{iso}}\lambda_c'\right) &= \frac{\partial \dot{G}_{\rm{iso}}}{\partial\lambda_c}\frac{\partial\lambda_c}{\partial \bar{{\rm{I}}}_1}\lambda_c'+ \dot{G}_{\rm{iso}}\frac{\partial\lambda_c'}{\partial\bar{{\rm{I}}}_1}\\
		&= \ddot{G}_{\rm{iso}}\lambda_c'^2+\dot{G}_{\rm{iso}}\lambda_c''.
		\label{eq:chain-4}
	\end{aligned}
\end{equation}
First and second derivatives of the indefinite integral with respect to the current stretch $\lambda_c(\bar{{\rm{I}}}_1)$ which is necessary for derivation of aforementioned derivatives of the indefinite integral with product rule are as follows:
\begin{equation}
	\begin{aligned}
		\dot{G}_{\rm{iso}}\left(\lambda_c(\bar{{\rm{I}}}_1)\right) &= \frac{^{\rm{iso}}a_0 ^{\rm{iso}}a_2}{\left(\lambda_c-1\right)}\rm{exp}\left(\frac{1}{4\, ^{\rm{iso}}a_1} - \frac{\left(1+2 ^{\rm{iso}}a_1 \rm{ln}\left(^{\rm{iso}}a_2 (\lambda_c-1) \right) \right)^2}{4\, ^{\rm{iso}}a_1}\right),\\
		\ddot{G}_{\rm{iso}}\left(\lambda_c(\bar{\rm{I}}_1)\right) &= \frac{^{\rm{iso}}a_0 ^{\rm{iso}}a_2}{\left(\lambda_c-1\right)^2}\rm{exp}\left(\frac{1}{4\, ^{\rm{iso}}a_1} - \frac{\left(1+2 ^{\rm{iso}}a_1 \rm{ln}\left(^{\rm{iso}}a_2 (\lambda_c-1) \right) \right)^2}{4\, ^{\rm{iso}}a_1}\right)\\
		&\hspace{25mm} \left(2+2 ^{\rm{iso}}a_1 \rm{ln}\left(^{\rm{iso}}a_2 (\lambda_c-1) \right) \right).
	\end{aligned}
	\label{eq:chain-5}
\end{equation}
First and second derivatives of the isochoric current chain stretch $\lambda_c\left(\bar{\rm{I}}_1 \right)$ with respect to $\bar{{\rm I}}_1$ follows
\begin{equation}
	\begin{aligned}
		\lambda_c'\left(\bar{\rm{I}}_1 \right) = \frac{1}{6}\left( \frac{\bar{\rm{I}}_1}{3}\right)^{-1/2}\!\!\!,\hspace{5mm}
		\lambda_c''\left(\bar{\rm{I}}_1 \right) = -\frac{1}{36}\left( \frac{\bar{\rm{I}}_1}{3}\right)^{-3/2}.
		\label{eq:chain-8}
	\end{aligned}
\end{equation}
The first and second order derivatives of the isochoric micromechanical free energy function are derived with the help of the aforementioned derivations of $G_{\rm{iso}}\left( \lambda_c (\bar{\rm{I}}_1)\right)$ and $\lambda_c(\bar{\rm{I}}_1)$. These derivatives of the isochoric free energy $W_{\rm{iso}}$ followed by
\begin{equation}
	\begin{aligned}
		W_{\rm{iso}}' &= G_{\rm{iso}}{\rm \Psi}_{\rm{iso}}' + G_{\rm{iso}}'{\rm \Psi}_{\rm{iso}} = G_{\rm{iso}} {\rm \Psi}_{\rm{iso}}' + \dot{G}_{\rm{iso}}\lambda_c'{\rm \Psi}_{\rm{iso}},\\
		W_{\rm{iso}}'' &= G_{\rm{iso}}{\rm \Psi}_{\rm{iso}}'' +2G_{\rm{iso}}'{\rm \Psi}_{\rm{iso}}' + G_{\rm{iso}}''{\rm \Psi}_{\rm{iso}}\\
		&= G_{\rm{iso}} {\rm \Psi}_{\rm{iso}}'' + 2\dot{G}_{\rm{iso}}\lambda_c'{\rm \Psi}_{\rm{iso}} + \left(\ddot{G}_{\rm{iso}}\lambda_c'^2 +\dot{G}_{\rm{iso}}\lambda_c'' \right){\rm \Psi}_{\rm{iso}}.
	\end{aligned}
	\label{eq:chain-6}
\end{equation}
The first and second order derivatives of the indefinite integral of the volumetric component are derived with the help of chain rule similar to the isochoric component. Due to the implicit dependence of the infinite integral on $\lambda_c(J)=J^{1/3}$, the derivative of the infinite integral must be derived with the chain rule. The first\footnote{First order derivatives of the components: $\frac{\partial(\bullet)}{\partial\lambda_c}=\dot{(\bullet)}$ and $\frac{\partial(\bullet)}{\partial J}=(\bullet)'$\\} derivative and second\footnote{Second order derivatives of the components: $\frac{\partial^2(\bullet)}{\partial\lambda_c^2}=\ddot{(\bullet)}$ and $\frac{\partial^2(\bullet)}{\partial J^2}=(\bullet)''$} derivatives are followed as
\begin{equation}
	\begin{aligned}
		G'_{\rm{vol}}\left(\lambda_c(J)\right) &= \frac{\partial G_{\rm{vol}}}{\partial\lambda_c}\frac{\partial\lambda_c}{\partial J} =\dot{G}_{\rm{vol}}\lambda_c',\\
		G''_{\rm{vol}}\left(\lambda_c(J)\right) &= \frac{\partial \dot{G}_{\rm{vol}}}{\partial\lambda_c}\frac{\partial\lambda_c}{\partial J}\lambda_c'+ \dot{G}_{\rm{vol}}\frac{\partial\lambda_c'}{\partial J}\\
		&= \ddot{G}_{\rm{vol}}\lambda_c'^2+\dot{G}_{\rm{vol}}\lambda_c''.
		\label{eq:chain-13}
	\end{aligned}
\end{equation}
Whereas the first and second derivatives of the indefinite integral with respect to $\lambda_c(J)$ in the aforementioned equation \eqref{eq:chain-13} are as follows
\begin{equation}
	\begin{aligned}
		\dot{G}_{\rm{vol}}\left(\lambda_c(J)\right) &= \frac{^{\rm{vol}}a_0 ^{\rm{vol}}a_2}{\left(\lambda_c-1\right)}\rm{exp}\left(\frac{1}{4\, ^{\rm{vol}}a_1} - \frac{\left(1+2 ^{\rm{vol}}a_1 \rm{ln}\left(^{\rm{vol}}a_2 (\lambda_c-1) \right) \right)^2}{4\, ^{\rm{vol}}a_1}\right),\\
		\ddot{G}_{\rm{vol}}\left(\lambda_c(J)\right) &= \frac{^{\rm{vol}}a_0 ^{\rm{vol}}a_2}{\left(\lambda_c-1\right)^2} \rm{exp}\left(\frac{1}{4\, ^{\rm{vol}}a_1} - \frac{\left(1+2 ^{\rm{vol}}a_1 \rm{ln}\left(^{\rm{vol}}a_2 (\lambda_c-1) \right) \right)^2}{4\, ^{\rm{vol}}a_1}\right)\\
		&\hspace{25mm}\left(2+2 ^{\rm{vol}}a_1 \rm{ln}\left(^{\rm{vol}}a_2 (\lambda_c-1) \right) \right).
	\end{aligned}
	\label{eq:chain-15}
\end{equation}
The first and second order derivative of $\lambda_c(J)$ with respect to $J$ are
\begin{equation}
	\begin{aligned}
		\lambda_c'\left(J \right) &= \frac{1}{3}{J}^{-1/3}; \hspace{5 mm}
		\lambda_c''\left(J \right) &= -\frac{1}{9}{J}^{-4/3}.
	\end{aligned}
	\label{eq:chain-17}
\end{equation}
The first and second derivatives of the volumetric $W_{\rm vol}$ component of the micromechanical free energy are necessary to derive stress and tangent tensors. These derivatives are defined with the help of previously mentioned derivations of $G_{\rm{vol}}\left(\lambda_c(J)\right)$ and $\lambda_c(J)$ as follows
\begin{equation}
	\begin{aligned}
		W_{\rm{vol}}' &= G_{\rm{vol}}{\rm \Psi}_{\rm{vol}}' + G_{\rm{vol}}'{\rm \Psi}_{\rm{vol}} = G_{\rm{vol}} {\rm \Psi}_{\rm{vol}}' + \dot{G}_{\rm{vol}}\lambda_c'{\rm \Psi}_{\rm{vol}},\\
		W_{\rm{vol}}'' &= G_{\rm{vol}}{\rm \Psi}_{\rm{vol}}'' +2G_{\rm{vol}}'{\rm \Psi}_{\rm{vol}}' + G_{\rm{vol}}''{\rm \Psi}_{\rm{vol}}\\
		&= G_{\rm{vol}} {\rm \Psi}_{\rm{vol}}'' + 2\dot{G}_{\rm{vol}}\lambda_c'{\rm \Psi}_{\rm{vol}} + \left(\ddot{G}_{\rm{vol}}\lambda_c'^2 +\dot{G}_{\rm{vol}}\lambda_c'' \right){\rm \Psi}_{\rm{vol}}
	\end{aligned}
	\label{eq:chain-16-2}
\end{equation}
For non-linear formulations that are often solved with the Newton increment method requires the evaluation of both the stress and tangent tensor. The measure of stresses in the current configuration can be expressed in the updated Lagrangian formulation with stress tensor $\mathbf{T}$ and the individual components of the stress defined in the equation \eqref{finvisko440} are evaluated with the equations \eqref{eq:totener} follows
\begin{equation}
	\begin{split}
		\mathbf{T}_{\rm eq}\left({\bar{\mathbf{B}}},\lambda_{m}\right) &= 2\mathbf{B}\cdot\frac{\partial W_{\rm eq} \left(\bar{\rm I}_1^{\bar{\mathbf{B}}},\lambda_{m} \right)}{\partial\mathbf{B}} = 2J^{-2/3}\left(W_{\rm eq}\right)'\left(\bar{\mathbf{B}} - \frac{1}{3}\bar{\rm I}_1^{\bar{\mathbf{B}}}\mathbf{I}\right);\\
		\mathbf{T}_{\rm vol}\left(J,\lambda_{m}\right) &= 2\mathbf{B}\cdot\frac{\partial W_{\rm vol} \left(J,\lambda_{m} \right)}{\partial\mathbf{B}} = J W_{\rm vol}\mathbf{I};\\
		\mathbf{T}^{j}_{\rm neq}\left({\bar{\mathbf{B}}^j_e},\lambda_{m}\right) &= 2\bar{\mathbf{B}}^j_e\cdot\frac{\partial W_{\rm neq}^j \left(\bar{\rm I}_1^{\bar{\mathbf{B}}^j_e},\lambda_{m} \right)}{\partial\bar{\mathbf{B}}^j_e} = 2J^{-2/3}\left(W_{\rm neq}^j\right)'\left(\bar{\mathbf{B}}^j_e - \frac{1}{3}\bar{\rm I}_1^{\bar{\mathbf{B}}^j_e}\mathbf{I}\right).
		\label{stressten}
	\end{split}
\end{equation}
The fourth order tangent tensor is derived as the second order derivative of the free energy density. This tensor is derived with the free energy function given in the equations \eqref{eq:totener} follows
\begin{equation}
	\begin{aligned}
		\boldsymbol{\kappa} &= 4\mathbf{B}\cdot\frac{\partial^2 W \left(\bar{\rm I}_1^{\bar{\mathbf{B}}}, \bar{\rm I}_1^{\bar{\mathbf{B}}^j_e},J,m \right)}{\partial\mathbf{B}\,\partial\mathbf{B}}\cdot\mathbf{B} =\,\,\, \boldsymbol{\kappa}{}_{\rm eq} + \boldsymbol{\kappa}{}_{\rm vol} + \boldsymbol{\kappa}{}^{j}_{\rm neq},\\
		\boldsymbol{\kappa}{}_{\rm eq} = 4\mathbf{B}\cdot&\frac{\partial^2 W_{\rm eq} \left(\bar{\rm I}_1^{\bar{\mathbf{B}}}, m \right)}{\partial\mathbf{B}\, \partial\mathbf{B}}\cdot\mathbf{B}; \hspace{5mm} \boldsymbol{\kappa}{}_{\rm vol} =4\mathbf{B}\cdot\frac{\partial^2 W_{\rm vol} \left(J,m \right)}{\partial\mathbf{B}\,\partial\mathbf{B}}\cdot\mathbf{B},\\
		\boldsymbol{\kappa}{}^{j}_{\rm neq} = 4\bar{\mathbf{B}}_e^j\cdot&\frac{\partial^2 W_{\rm neq}^j \left(\bar{\rm I}_1^{\bar{\mathbf{B}}^j_e},m \right)}{\partial\bar{\mathbf{B}}_e^j\, \partial\bar{\mathbf{B}}_e^j}\cdot\bar{\mathbf{B}}_e^j,
	\end{aligned}
\end{equation}
where the individual components of the tangent tensor are evaluated as 
\begin{equation}
	\begin{aligned}
		\boldsymbol{\kappa}_{\rm eq} =\,& 4 \left(W_{\rm eq} \left(\bar{\rm I}_1^{\bar{\mathbf{B}}},m\right)\right)' \left( \mathbf{I}\otimes\bar{\mathbf{B}}\right)^{s_{24}} + 4 \left(W_{\rm eq} \left(\bar{\rm I}_1^{\bar{\mathbf{B}}},m\right)\right)'' \left(\bar{\mathbf{B}}\otimes\bar{\mathbf{B}}\right) - \\
		&\frac{4}{3}\left(\bar{\rm I}_1^{\bar{\mathbf{B}}} \left(W_{\rm eq} \left(\bar{\rm I}_1^{\bar{\mathbf{B}}},m\right)\right)''\!\! + \left(W_{\rm eq} \left(\bar{\rm I}_1^{\bar{\mathbf{B}}},m \right) \right)' \right) \left(\bar{\mathbf{B}}\otimes\mathbf{I}+\mathbf{I}\otimes\bar{\mathbf{B}}\right)+\\
		&\frac{4}{9}\left(\left(\bar{\rm I}_1^{\bar{\mathbf{B}}}\right)^2 \left(W_{\rm eq} \left(\bar{\rm I}_1^{\bar{\mathbf{B}}},m\right)\right)''\!\! + \bar{\rm I}_1^{\bar{\mathbf{B}}} \left(W_{\rm eq} \left(\bar{\rm I}_1^{\bar{\mathbf{B}}},m\right)\right)' \right)\left(\mathbf{I}\otimes\mathbf{I}\right)\\
		\boldsymbol{\kappa}_{\rm vol} =\,& \left(J^2 \left(W_{\rm vol} \left(J,m \right)\right)'' + J \left(W_{\rm vol} \left(J,m \right)\right)' \right)\mathbf{I}\otimes\mathbf{I}\\
		\boldsymbol{\kappa}^{j}_{\rm neq} =\,& 4 W'_{\rm neq} \left(\bar{\rm I}_1^{\bar{\mathbf{B}}^j_e}, m \right) \left(\mathbf{I} \otimes \bar{\mathbf{B}}\right)^{s_{24}} + 4 W''_{\rm neq} \left(\bar{\rm I}_1^{\bar{\mathbf{B}}^j_e}, m \right) \left( \bar{\mathbf{B}} \otimes\bar{\mathbf{B}}\right) - \\
		&\frac{4}{3}\left(\bar{\rm I}_1^{\bar{\mathbf{B}}^j_e} W''_{\rm neq} \left(\bar{\rm I}_1^{\bar{\mathbf{B}}^j_e}, m \right) + W'_{\rm neq} \left(\bar{\rm I}_1^{\bar{\mathbf{B}}^j_e}, m \right) \right)\left(\bar{\mathbf{B}}\otimes\mathbf{I}+\mathbf{I}\otimes\bar{\mathbf{B}}\right)+\\
		&\frac{4}{9}\left(\left(\bar{\rm I}_1^{\bar{\mathbf{B}}^j_e}\right)^2 W''_{\rm neq} \left(\bar{\rm I}_1^{\bar{\mathbf{B}}^j_e},m \right) + \bar{\rm I}_1^{\bar{\mathbf{B}}^j_e} W'_{\rm neq} \left(\bar{\rm I}_1^{\bar{\mathbf{B}}^j_e},m \right)\right)\left(\mathbf{I}\otimes\mathbf{I}\right).
	\end{aligned}
	\label{tangent}
\end{equation}
The free energy functions required in implementing the viscoelastic material model is motivated by the micromechanical free energy function defined the equation \eqref{micromech}. In which the isochoric component of the equilibrium free energy $W_{\rm eq}$ and the non-equilibrium free energy $W^j_{\rm neq}$ fo $j^{th}$ Maxwell element are introduced with the isochoric component of the micromechanical free energy defined in the equation \eqref{eq:chain-1}. The micromechanical free energy function is based on the multiplicative relation between the free energy function and probable chain length distribution, in which the free energy function is defined with Neo-Hook free energy. To this end, the micromechanical free energy functions takes the form
\begin{equation}
	\begin{aligned}
		&{\rm \Psi}_{\rm eq}\left(\bar{{\rm I}}_1^{\bar{\mathbf{B}}}\right)=c_{10}\left(\bar{{\rm I}}_1^{\bar{\mathbf{B}}}-3\right) {\rm results \, in}\, W_{\rm eq} = {\rm \Psi}_{\rm eq}\left(\bar{{\rm I}}_1^{\bar{\mathbf{B}}}\right)G\left(\lambda_c\left(\bar{{\rm I}}_1^{\bar{\mathbf{B}}}\right)\right),\\
		&{\rm \Psi}^j_{\rm neq}\left(\bar{{\rm I}}_1^{\bar{\mathbf{B}}_e^j}\right)=c_{10j}\left(\bar{{\rm I}}_1^{\bar{\mathbf{B}}_e^j}-3\right) {\rm results \, in}\, W^j_{\rm neq} = {\rm \Psi}_{\rm eq}\left(\bar{{\rm I}}_1^{\bar{\mathbf{B}}_e^j}\right)G\left(\lambda_c\left(\bar{{\rm I}}_1^{\bar{\mathbf{B}}_e^j}\right)\right),
	\end{aligned}
\end{equation}
where, first and second order derivatives of the free energy required to evaluate the stress and tangent tensors follow the equation \eqref{eq:chain-6}. Whereas the free energy function required in formulation of volumetric components of the micromechanical free energy considers the quadratic function of Jacobian
\begin{equation}
	{\rm \Psi}_{\rm vol}\left(J\right)= \frac{1}{D}\left(J - 1\right)^2 {\rm results \, in}\, W_{\rm vol} = {\rm \Psi}_{\rm vol}\left(J\right)G\left(\lambda_c\left(J\right)\right),
\end{equation}
where, $D$ corresponds to the shear modulus. When $J=1$ the material corresponds to the incompressible material. The first and second order derivatives of the volumetric free energy are evaluated as discussed in the equation \eqref{eq:chain-16-2}. As result of the additive decomposition of the equilibrium component of the free energy function into isochoric and volumetric components. The stress tensor $\mathbf{T}$ is introduced as sum of the individual stress components 
\begin{equation}
	\mathbf{T}\,=\,\mathbf{T}_{\rm vol}(J,\lambda_m) + \mathbf{T}_{\rm eq}(\bar{\mathbf{B}},\lambda_m) + \sum_{j=1}^n\mathbf{T}_{\rm neq}^j(\bar{\mathbf{B}}_e^j,\lambda_m),
	\label{finvisko440}
\end{equation}
where, $\mathbf{T}_{\rm eq}$, $\mathbf{T}^{\rm eq}_{\rm vol}$ and $\mathbf{T}_{\rm neq}^j$ corresponds to the isochoric equilibrium stress component, volumetric stress component and the non-equilibrium stress component of the $j^{th}$ Maxwell element. 
\end{document}

%% file: defs.tex
\usepackage{amsfonts}
\usepackage{amssymb}
\usepackage{amsbsy}
\usepackage{ifthen}
\usepackage{multirow}
\usepackage{rotating}
\usepackage{color}
\usepackage{todonotes}
\usepackage{psfrag}
\usepackage{subcaption}
\usepackage{graphicx}

\usepackage{graphicx}
\usepackage[parfill]{parskip}
\usepackage{float}
\usepackage{amssymb, amsmath}
\usepackage{color, fancyhdr}
\usepackage{latexsym, textcomp}
\usepackage{amssymb, amsmath}
\usepackage{lscape}
\usepackage{ae}
\usepackage[T1]{fontenc}
\usepackage{tabularx}
\usepackage[mathscr]{eucal}
\usepackage{longtable}
\usepackage{pstricks}
\usepackage{pst-grad,pst-3d}
\usepackage{color}
\usepackage{enumitem}
\usepackage{epsf,exscale,psfrag,graphics}
\usepackage{xfrac}
\usepackage{placeins}
\usepackage{multirow}

\newcommand{\beq}{\begin{equation}}

\newcommand{\eeq}[1]{%
  \ifthenelse{\equal{#1}{ }}
  {
    \end{equation}
  }
  {
    \label{eq:#1}
    \end{equation}
  }
}

\newcommand{\beqn}      {\begin{eqnarray}}
\newcommand{\eeqn}      {\end{eqnarray}}
\newcommand     {\bea}          {\begin{equationarray}}
\newcommand     {\eea}          {\end{equationarray}}
\newcommand{\beqa}{\begin{eqnarray}}
\newcommand{\eeqa}{\end{eqnarray}}

\newcommand{\bit}       {\begin{itemize}}
\newcommand{\eit}       {\end{itemize}}

\newcommand{\bde}       {\begin{description}}
\newcommand{\ede}       {\end{description}}

\newcommand{\bec}       {\begin{center}}
\newcommand{\eec}       {\end{center}}

\newsavebox{\BildText}
                        {\centerline{\usebox{\BildText}}\end{figure}}
                        
\newcommand{\Kasten}[3]%
{\begin{equation}
  \fbox{
    \begin{minipage}{5.5in}
      \centering
      #2
      \rule{5.4in}{0.5pt}
      {\bf #3}
    \end{minipage}
  }
  \label{box:#1}
\end{equation}}

\newcommand{\RefBraces}[1]%
{%
  \typeout{Referenz #1 in Seite \thepage}%
  \ifthenelse{\equal{\pageref{#1}}{\thepage}}%
             {(\ref{#1})}%
             {(\ref{#1}) auf Seite \pageref{#1}}%
}

\DeclareMathAlphabet\eufm{U}{euf}{m}{n}
\DeclareMathAlphabet\eufb{U}{euf}{b}{n}
\DeclareMathAlphabet\eurm{U}{eur}{m}{n}
\DeclareMathAlphabet\eurb{U}{eur}{b}{n}
\DeclareMathAlphabet\eusm{U}{eus}{m}{n}
\DeclareMathAlphabet\eusb{U}{eus}{b}{n}
\DeclareMathAlphabet\mathsfm{OT1}{cmss}{m}{n}
\DeclareMathAlphabet\mathsfb{OT1}{cmss}{bx}{n}

\newcommand{\Qed}{\hfill \mbox{$\Box$}}         

\newcommand{\bigo}[1]{\setbox0\hbox{$\bigcirc$}
             \rlap{\raise .2ex\hbox to \wd0{\hfil ${\scriptscriptstyle
                   #1}$\hfil}}\box0}
 
\newcommand{\subf}[2]{%
	{\small\begin{tabular}[t]{l@{}c@{}}
			#1\\#2
	\end{tabular}}%
}
\newcommand{\svektor}[2]%
{\left(\!\!\!\begin{array}{c} #1\\#2 \end{array}\!\!\!\right)}

\newcommand{\zvektor}[2]%
{\left(#1,#2\right)^T}

\newcommand{\Svektor}[2]%
{\left[\!\!\!\begin{array}{l} #1\\#2 \end{array}\!\!\!\right]}

\newcommand{\Zvektor}[2]%
{\left[#1,#2\right]^T}

%% file: tension/rheologisches_modell1.eps_tex
\begingroup%
  \makeatletter%
  \providecommand\color[2][]{%
    \errmessage{(Inkscape) Color is used for the text in Inkscape, but the package 'color.sty' is not loaded}%
    \renewcommand\color[2][]{}%
  }%
  \providecommand\transparent[1]{%
    \errmessage{(Inkscape) Transparency is used (non-zero) for the text in Inkscape, but the package 'transparent.sty' is not loaded}%
    \renewcommand\transparent[1]{}%
  }%
  \providecommand\rotatebox[2]{#2}%
  \ifx\svgwidth\undefined%
    \setlength{\unitlength}{1376.00006104bp}%
    \ifx\svgscale\undefined%
      \relax%
    \else%
      \setlength{\unitlength}{\unitlength * \real{\svgscale}}%
    \fi%
  \else%
    \setlength{\unitlength}{\svgwidth}%
  \fi%
  \global\let\svgwidth\undefined%
  \global\let\svgscale\undefined%
  \makeatother%
  \begin{picture}(1,0.42877906)%
    \put(0,0){\includegraphics[width=\unitlength]{tension/rheologisches_modell1.eps}}%
    \put(-0.05550893,0.20363679){\color[rgb]{0,0,0}\makebox(0,0)[lb]{\smash{$\mu_{1}$}}}%
    \put(0.16443296,0.27494563){\color[rgb]{0,0,0}\makebox(0,0)[lb]{\smash{$\mu_{11}$}}}%
    \put(0.422265,0.27494563){\color[rgb]{0,0,0}\makebox(0,0)[lb]{\smash{$\mu_{12}$}}}%
    \put(0.77456092,0.27494563){\color[rgb]{0,0,0}\makebox(0,0)[lb]{\smash{$\mu_{1n}$}}}%
    \put(0.17642424,0.1059776){\color[rgb]{0,0,0}\makebox(0,0)[lb]{\smash{$\eta_1$}}}%
    \put(0.44193518,0.1059776){\color[rgb]{0,0,0}\makebox(0,0)[lb]{\smash{$\eta_2$}}}%
    \put(0.78804765,0.1059776){\color[rgb]{0,0,0}\makebox(0,0)[lb]{\smash{$\eta_n$}}}%
  \end{picture}%
\endgroup%

%% file: tension/2d.eps_tex
\begingroup%
  \makeatletter%
  \providecommand\color[2][]{%
    \errmessage{(Inkscape) Color is used for the text in Inkscape, but the package 'color.sty' is not loaded}%
    \renewcommand\color[2][]{}%
  }%
  \providecommand\transparent[1]{%
    \errmessage{(Inkscape) Transparency is used (non-zero) for the text in Inkscape, but the package 'transparent.sty' is not loaded}%
    \renewcommand\transparent[1]{}%
  }%
  \providecommand\rotatebox[2]{#2}%
  \ifx\svgwidth\undefined%
    \setlength{\unitlength}{621bp}%
    \ifx\svgscale\undefined%
      \relax%
    \else%
      \setlength{\unitlength}{\unitlength * \real{\svgscale}}%
    \fi%
  \else%
    \setlength{\unitlength}{\svgwidth}%
  \fi%
  \global\let\svgwidth\undefined%
  \global\let\svgscale\undefined%
  \makeatother%
  \begin{picture}(1,0.33494365)%
    \put(0,0){\includegraphics[width=\unitlength]{tension/2d.eps}}%
    \put(0.37,0.20288266){\color[rgb]{0,0,0}\makebox(0,0)[lb]{\smash{$\rm A$}}}%
    \put(0.35,0.29517713){\color[rgb]{0,0,0}\makebox(0,0)[lb]{\smash{$\rm 75$}}}%
    \put(0.92,0.12252653){\color[rgb]{0,0,0}\makebox(0,0)[lb]{\smash{$\rm 12.5$}}}%
    \put(0.86,0.12252653){\color[rgb]{0,0,0}\makebox(0,0)[lb]{\smash{$\rm 4$}}}%
    \put(0.67,0.23520259){\color[rgb]{0,0,0}\makebox(0,0)[lb]{\smash{$\rm 15.43$}}}%
    \put(0.25,0.21970961){\color[rgb]{0,0,0}\makebox(0,0)[lb]{\smash{$\rm R60$}}}%
    \put(0.37,0.02780821){\color[rgb]{0,0,0}\makebox(0,0)[lb]{\smash{$\rm A$}}}%
  \end{picture}%
\endgroup%

%% file: tension/BF5e4_EXPVSSIM_0.tex
\begingroup
\vspace{10mm}
\large
  \makeatletter
  \providecommand\color[2][]{%
    \GenericError{(gnuplot) \space\space\space\@spaces}{%
      Package color not loaded in conjunction with
      terminal option `colourtext'%
    }{See the gnuplot documentation for explanation.%
    }{Either use 'blacktext' in gnuplot or load the package
      color.sty in LaTeX.}%
    \renewcommand\color[2][]{}%
  }%
  \providecommand\includegraphics[2][]{%
    \GenericError{(gnuplot) \space\space\space\@spaces}{%
      Package graphicx or graphics not loaded%
    }{See the gnuplot documentation for explanation.%
    }{The gnuplot epslatex terminal needs graphicx.sty or graphics.sty.}%
    \renewcommand\includegraphics[2][]{}%
  }%
  \providecommand\rotatebox[2]{#2}%
  \@ifundefined{ifGPcolor}{%
    \newif\ifGPcolor
    \GPcolortrue
  }{}%
  \@ifundefined{ifGPblacktext}{%
    \newif\ifGPblacktext
    \GPblacktextfalse
  }{}%
  \let\gplgaddtomacro\g@addto@macro
  \gdef\gplbacktext{}%
  \gdef\gplfronttext{}%
  \makeatother
  \ifGPblacktext
    \def\colorrgb#1{}%
    \def\colorgray#1{}%
  \else
    \ifGPcolor
      \def\colorrgb#1{\color[rgb]{#1}}%
      \def\colorgray#1{\color[gray]{#1}}%
      \expandafter\def\csname LTw\endcsname{\color{white}}%
      \expandafter\def\csname LTb\endcsname{\color{black}}%
      \expandafter\def\csname LTa\endcsname{\color{black}}%
      \expandafter\def\csname LT0\endcsname{\color[rgb]{1,0,0}}%
      \expandafter\def\csname LT1\endcsname{\color[rgb]{0,1,0}}%
      \expandafter\def\csname LT2\endcsname{\color[rgb]{0,0,1}}%
      \expandafter\def\csname LT3\endcsname{\color[rgb]{1,0,1}}%
      \expandafter\def\csname LT4\endcsname{\color[rgb]{0,1,1}}%
      \expandafter\def\csname LT5\endcsname{\color[rgb]{1,1,0}}%
      \expandafter\def\csname LT6\endcsname{\color[rgb]{0,0,0}}%
      \expandafter\def\csname LT7\endcsname{\color[rgb]{1,0.3,0}}%
      \expandafter\def\csname LT8\endcsname{\color[rgb]{0.5,0.5,0.5}}%
    \else
      \def\colorrgb#1{\color{black}}%
      \def\colorgray#1{\color[gray]{#1}}%
      \expandafter\def\csname LTw\endcsname{\color{white}}%
      \expandafter\def\csname LTb\endcsname{\color{black}}%
      \expandafter\def\csname LTa\endcsname{\color{black}}%
      \expandafter\def\csname LT0\endcsname{\color{black}}%
      \expandafter\def\csname LT1\endcsname{\color{black}}%
      \expandafter\def\csname LT2\endcsname{\color{black}}%
      \expandafter\def\csname LT3\endcsname{\color{black}}%
      \expandafter\def\csname LT4\endcsname{\color{black}}%
      \expandafter\def\csname LT5\endcsname{\color{black}}%
      \expandafter\def\csname LT6\endcsname{\color{black}}%
      \expandafter\def\csname LT7\endcsname{\color{black}}%
      \expandafter\def\csname LT8\endcsname{\color{black}}%
    \fi
  \fi
  \setlength{\unitlength}{0.0500bp}%
  \begin{picture}(7200.00,5040.00)%
    \gplgaddtomacro\gplbacktext{%
      \csname LTb\endcsname%
      \put(682,704){\makebox(0,0)[r]{\strut{} 0}}%
      \put(682,1383){\makebox(0,0)[r]{\strut{} 2}}%
      \put(682,2061){\makebox(0,0)[r]{\strut{} 4}}%
      \put(682,2740){\makebox(0,0)[r]{\strut{} 6}}%
      \put(682,3418){\makebox(0,0)[r]{\strut{} 8}}%
      \put(682,4097){\makebox(0,0)[r]{\strut{} 10}}%
      \put(682,4775){\makebox(0,0)[r]{\strut{} 12}}%
      \put(814,484){\makebox(0,0){\strut{} 1}}%
      \put(1670,484){\makebox(0,0){\strut{} 1.1}}%
      \put(2525,484){\makebox(0,0){\strut{} 1.2}}%
      \put(3381,484){\makebox(0,0){\strut{} 1.3}}%
      \put(4236,484){\makebox(0,0){\strut{} 1.4}}%
      \put(5092,484){\makebox(0,0){\strut{} 1.5}}%
      \put(5947,484){\makebox(0,0){\strut{} 1.6}}%
      \put(6803,484){\makebox(0,0){\strut{} 1.7}}%
      \put(176,2739){\rotatebox{-270}{\makebox(0,0){\strut{}Stress $\left[{\rm MPa}\right]$}}}%
      \put(3940,154){\makebox(0,0){\strut{}Stretch $\left[-\right]$}}%
    }%
    \gplgaddtomacro\gplfronttext{%
      \csname LTb\endcsname%
      \put(3322,4602){\makebox(0,0)[r]{\strut{}Standard deviation}}%
      \csname LTb\endcsname%
      \put(3322,4382){\makebox(0,0)[r]{\strut{}$0\%$ r.H. Experiment}}%
      \csname LTb\endcsname%
      \put(3322,4162){\makebox(0,0)[r]{\strut{}$0\%$ r.H. Simulation}}%
    }%
    \gplgaddtomacro\gplbacktext{%
      \csname LTb\endcsname%
      \put(682,704){\makebox(0,0)[r]{\strut{} 0}}%
      \put(682,1383){\makebox(0,0)[r]{\strut{} 2}}%
      \put(682,2061){\makebox(0,0)[r]{\strut{} 4}}%
      \put(682,2740){\makebox(0,0)[r]{\strut{} 6}}%
      \put(682,3418){\makebox(0,0)[r]{\strut{} 8}}%
      \put(682,4097){\makebox(0,0)[r]{\strut{} 10}}%
      \put(682,4775){\makebox(0,0)[r]{\strut{} 12}}%
      \put(814,484){\makebox(0,0){\strut{} 1}}%
      \put(1670,484){\makebox(0,0){\strut{} 1.1}}%
      \put(2525,484){\makebox(0,0){\strut{} 1.2}}%
      \put(3381,484){\makebox(0,0){\strut{} 1.3}}%
      \put(4236,484){\makebox(0,0){\strut{} 1.4}}%
      \put(5092,484){\makebox(0,0){\strut{} 1.5}}%
      \put(5947,484){\makebox(0,0){\strut{} 1.6}}%
      \put(6803,484){\makebox(0,0){\strut{} 1.7}}%
      \put(176,2739){\rotatebox{-270}{\makebox(0,0){\strut{}Stress $\left[{\rm MPa}\right]$}}}%
      \put(3940,154){\makebox(0,0){\strut{}Stretch $\left[-\right]$}}%
    }%
    \gplgaddtomacro\gplfronttext{%
      \csname LTb\endcsname%
      \put(3322,4602){\makebox(0,0)[r]{\strut{}Standard deviation}}%
      \csname LTb\endcsname%
      \put(3322,4382){\makebox(0,0)[r]{\strut{}$0\%$ r.H. Experiment}}%
      \csname LTb\endcsname%
      \put(3322,4162){\makebox(0,0)[r]{\strut{}$0\%$ r.H. Simulation}}%
    }%
    \gplbacktext
    \put(0,0){\includegraphics{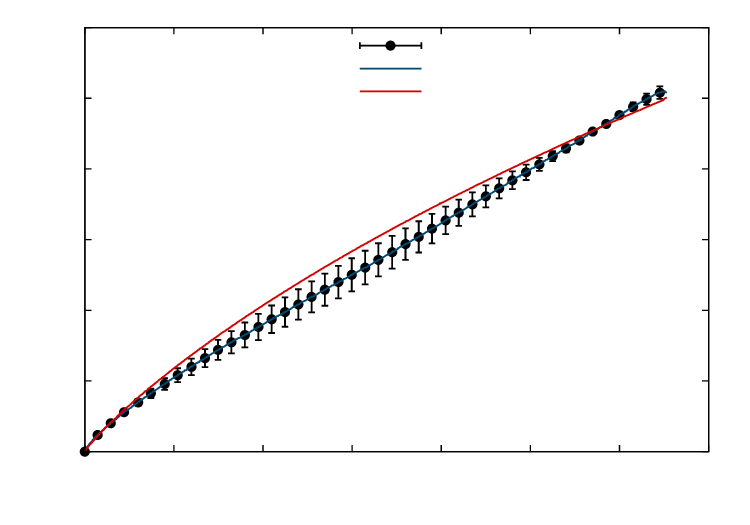}}%
    \gplfronttext
  \end{picture}%
\endgroup

%% file: tension/angular_spec.eps_tex
\begingroup%
\vspace{-1cm}
  \makeatletter%
  \providecommand\color[2][]{%
    \errmessage{(Inkscape) Color is used for the text in Inkscape, but the package 'color.sty' is not loaded}%
    \renewcommand\color[2][]{}%
  }%
  \providecommand\transparent[1]{%
    \errmessage{(Inkscape) Transparency is used (non-zero) for the text in Inkscape, but the package 'transparent.sty' is not loaded}%
    \renewcommand\transparent[1]{}%
  }%
  \providecommand\rotatebox[2]{#2}%
  \newcommand*\fsize{\dimexpr\f@size pt\relax}%
  \newcommand*\lineheight[1]{\fontsize{\fsize}{#1\fsize}\selectfont}%
  \ifx\svgwidth\undefined%
    \setlength{\unitlength}{657.63779528bp}%
    \ifx\svgscale\undefined%
      \relax%
    \else%
      \setlength{\unitlength}{\unitlength * \real{\svgscale}}%
    \fi%
  \else%
    \setlength{\unitlength}{\svgwidth}%
  \fi%
  \global\let\svgwidth\undefined%
  \global\let\svgscale\undefined%
  \makeatother%
  \begin{picture}(1,0.67672414)%
    \lineheight{1}%
    \setlength\tabcolsep{0pt}%
    \put(0,0){\includegraphics[width=\unitlength]{tension/angular_spec.eps}}%
    \put(0.41098363,0.58607728){\color[rgb]{1,1,1}\makebox(0,0)[lt]{\lineheight{1.25}\smash{\begin{tabular}[t]{l}$100$\end{tabular}}}}%
    \put(0.42,0.45908071){\color[rgb]{0,0,0}\makebox(0,0)[lt]{\lineheight{1.25}\smash{\begin{tabular}[t]{l}$90^\circ \pm 0.5^\circ$\end{tabular}}}}%
    \put(0.78,0.33){\color[rgb]{0,0,0}\makebox(0,0)[lt]{\lineheight{1.25}\smash{\begin{tabular}[t]{l}$\approx 22$\end{tabular}}}}%
    \put(0.22,0.28){\color[rgb]{0,0,0}\rotatebox{63.77423155}{\makebox(0,0)[lt]{\lineheight{1.25}\smash{\begin{tabular}[t]{l}${\rm R}19 \pm 0.05$\end{tabular}}}}}%
    \put(0.96,0.29){\color[rgb]{0,0,0}\rotatebox{88.54781331}{\makebox(0,0)[lt]{\lineheight{1.25}\smash{\begin{tabular}[t]{l}$19 \pm 0.05$\end{tabular}}}}}%
    \put(0.26,0.06){\color[rgb]{0,0,0}\rotatebox{47.20994265}{\makebox(0,0)[lt]{\lineheight{1.25}\smash{\begin{tabular}[t]{l}${\rm R}12.7 \pm 0.05$\end{tabular}}}}}%
    \put(0.63,0.21){\color[rgb]{0,0,0}\rotatebox{-55.50078137}{\makebox(0,0)[lt]{\lineheight{1.25}\smash{\begin{tabular}[t]{l}${\rm R}25.4 \pm 0.05$\end{tabular}}}}}%
  \end{picture}%
\endgroup%

%% file: tension/bcs1.eps_tex
\begingroup%
\vspace{15mm}
\large
  \makeatletter%
  \providecommand\color[2][]{%
    \errmessage{(Inkscape) Color is used for the text in Inkscape, but the package 'color.sty' is not loaded}%
    \renewcommand\color[2][]{}%
  }%
  \providecommand\transparent[1]{%
    \errmessage{(Inkscape) Transparency is used (non-zero) for the text in Inkscape, but the package 'transparent.sty' is not loaded}%
    \renewcommand\transparent[1]{}%
  }%
  \providecommand\rotatebox[2]{#2}%
  \ifx\svgwidth\undefined%
    \setlength{\unitlength}{712.5bp}%
    \ifx\svgscale\undefined%
      \relax%
    \else%
      \setlength{\unitlength}{\unitlength * \real{\svgscale}}%
    \fi%
  \else%
    \setlength{\unitlength}{\svgwidth}%
  \fi%
  \global\let\svgwidth\undefined%
  \global\let\svgscale\undefined%
  \makeatother%
  \begin{picture}(1,0.4)%
    \put(0,0){\includegraphics[width=\unitlength]{tension/bcs1.eps}}%
    \put(-0.02,0.32){\color[rgb]{0,0,0}\makebox(0,0)[lb]{\smash{$u_y$}}}%
    \put(1.,0.32){\color[rgb]{0,0,0}\makebox(0,0)[lb]{\smash{$u_y$}}}%
  \end{picture}%
\endgroup%

%% file: tension/contor1.eps_tex
\begingroup%
  \makeatletter%
  \providecommand\color[2][]{%
    \errmessage{(Inkscape) Color is used for the text in Inkscape, but the package 'color.sty' is not loaded}%
    \renewcommand\color[2][]{}%
  }%
  \providecommand\transparent[1]{%
    \errmessage{(Inkscape) Transparency is used (non-zero) for the text in Inkscape, but the package 'transparent.sty' is not loaded}%
    \renewcommand\transparent[1]{}%
  }%
  \providecommand\rotatebox[2]{#2}%
  \ifx\svgwidth\undefined%
    \setlength{\unitlength}{75bp}%
    \ifx\svgscale\undefined%
      \relax%
    \else%
      \setlength{\unitlength}{\unitlength * \real{\svgscale}}%
    \fi%
  \else%
    \setlength{\unitlength}{\svgwidth}%
  \fi%
  \global\let\svgwidth\undefined%
  \global\let\svgscale\undefined%
  \makeatother%
  \begin{picture}(1,3.56)%
    \put(0,-2.7){\includegraphics[width=\unitlength]{tension/contor1.eps}}%
    \put(1.3,5.2){\color[rgb]{0,0,0}\makebox(0,0)[lb]{\smash{0.75}}}%
    \put(-1.86296606,0.15084751){\color[rgb]{0,0,0}\makebox(0,0)[lb]{\smash{}}}%
    \put(1.3,7.8){\color[rgb]{0,0,0}\makebox(0,0)[lb]{\smash{1.00}}}%
    \put(1.3,2.6){\color[rgb]{0,0,0}\makebox(0,0)[lb]{\smash{0.50}}}%
    \put(1.3,0.0){\color[rgb]{0,0,0}\makebox(0,0)[lb]{\smash{0.25}}}%
    \put(1.3,-2.6){\color[rgb]{0,0,0}\makebox(0,0)[lb]{\smash{0.00}}}%
    \put(0.4,8.3){\color[rgb]{0,0,0}\makebox(0,0)[lb]{\smash{$\phi$}}}%
  \end{picture}%
\endgroup%

%% file: tension/crack2.tex
\begingroup
\large
  \makeatletter
  \providecommand\color[2][]{%
    \GenericError{(gnuplot) \space\space\space\@spaces}{%
      Package color not loaded in conjunction with
      terminal option `colourtext'%
    }{See the gnuplot documentation for explanation.%
    }{Either use 'blacktext' in gnuplot or load the package
      color.sty in LaTeX.}%
    \renewcommand\color[2][]{}%
  }%
  \providecommand\includegraphics[2][]{%
    \GenericError{(gnuplot) \space\space\space\@spaces}{%
      Package graphicx or graphics not loaded%
    }{See the gnuplot documentation for explanation.%
    }{The gnuplot epslatex terminal needs graphicx.sty or graphics.sty.}%
    \renewcommand\includegraphics[2][]{}%
  }%
  \providecommand\rotatebox[2]{#2}%
  \@ifundefined{ifGPcolor}{%
    \newif\ifGPcolor
    \GPcolortrue
  }{}%
  \@ifundefined{ifGPblacktext}{%
    \newif\ifGPblacktext
    \GPblacktextfalse
  }{}%
  \let\gplgaddtomacro\g@addto@macro
  \gdef\gplbacktext{}%
  \gdef\gplfronttext{}%
  \makeatother
  \ifGPblacktext
    \def\colorrgb#1{}%
    \def\colorgray#1{}%
  \else
    \ifGPcolor
      \def\colorrgb#1{\color[rgb]{#1}}%
      \def\colorgray#1{\color[gray]{#1}}%
      \expandafter\def\csname LTw\endcsname{\color{white}}%
      \expandafter\def\csname LTb\endcsname{\color{black}}%
      \expandafter\def\csname LTa\endcsname{\color{black}}%
      \expandafter\def\csname LT0\endcsname{\color[rgb]{1,0,0}}%
      \expandafter\def\csname LT1\endcsname{\color[rgb]{0,1,0}}%
      \expandafter\def\csname LT2\endcsname{\color[rgb]{0,0,1}}%
      \expandafter\def\csname LT3\endcsname{\color[rgb]{1,0,1}}%
      \expandafter\def\csname LT4\endcsname{\color[rgb]{0,1,1}}%
      \expandafter\def\csname LT5\endcsname{\color[rgb]{1,1,0}}%
      \expandafter\def\csname LT6\endcsname{\color[rgb]{0,0,0}}%
      \expandafter\def\csname LT7\endcsname{\color[rgb]{1,0.3,0}}%
      \expandafter\def\csname LT8\endcsname{\color[rgb]{0.5,0.5,0.5}}%
    \else
      \def\colorrgb#1{\color{black}}%
      \def\colorgray#1{\color[gray]{#1}}%
      \expandafter\def\csname LTw\endcsname{\color{white}}%
      \expandafter\def\csname LTb\endcsname{\color{black}}%
      \expandafter\def\csname LTa\endcsname{\color{black}}%
      \expandafter\def\csname LT0\endcsname{\color{black}}%
      \expandafter\def\csname LT1\endcsname{\color{black}}%
      \expandafter\def\csname LT2\endcsname{\color{black}}%
      \expandafter\def\csname LT3\endcsname{\color{black}}%
      \expandafter\def\csname LT4\endcsname{\color{black}}%
      \expandafter\def\csname LT5\endcsname{\color{black}}%
      \expandafter\def\csname LT6\endcsname{\color{black}}%
      \expandafter\def\csname LT7\endcsname{\color{black}}%
      \expandafter\def\csname LT8\endcsname{\color{black}}%
    \fi
  \fi
  \setlength{\unitlength}{0.0500bp}%
  \begin{picture}(7200.00,5040.00)%
    \gplgaddtomacro\gplbacktext{%
      \csname LTb\endcsname%
      \put(682,704){\makebox(0,0)[r]{\strut{} 0}}%
      \put(682,1518){\makebox(0,0)[r]{\strut{} 5}}%
      \put(682,2332){\makebox(0,0)[r]{\strut{} 10}}%
      \put(682,3147){\makebox(0,0)[r]{\strut{} 15}}%
      \put(682,3961){\makebox(0,0)[r]{\strut{} 20}}%
      \put(682,4775){\makebox(0,0)[r]{\strut{} 25}}%
      \put(814,484){\makebox(0,0){\strut{} 0}}%
      \put(1479,484){\makebox(0,0){\strut{} 5}}%
      \put(2145,484){\makebox(0,0){\strut{} 10}}%
      \put(2810,484){\makebox(0,0){\strut{} 15}}%
      \put(3476,484){\makebox(0,0){\strut{} 20}}%
      \put(4141,484){\makebox(0,0){\strut{} 25}}%
      \put(4807,484){\makebox(0,0){\strut{} 30}}%
      \put(5472,484){\makebox(0,0){\strut{} 35}}%
      \put(6138,484){\makebox(0,0){\strut{} 40}}%
      \put(6803,484){\makebox(0,0){\strut{} 45}}%
      \put(176,2739){\rotatebox{-270}{\makebox(0,0){\strut{}Force $\left[\rm N\right]$}}}%
      \put(4072,154){\makebox(0,0){\strut{}Time $\left[\rm s\right]$}}%
    }%
    \gplgaddtomacro\gplfronttext{%
      \csname LTb\endcsname%
      \put(3322,4514){\makebox(0,0)[r]{\strut{}Standard deviation}}%
      \csname LTb\endcsname%
      \put(3322,4118){\makebox(0,0)[r]{\strut{}Experiment}}%
      \csname LTb\endcsname%
      \put(3322,3722){\makebox(0,0)[r]{\strut{}Simulation}}%
    }%
    \gplbacktext
    \put(0,0){\includegraphics{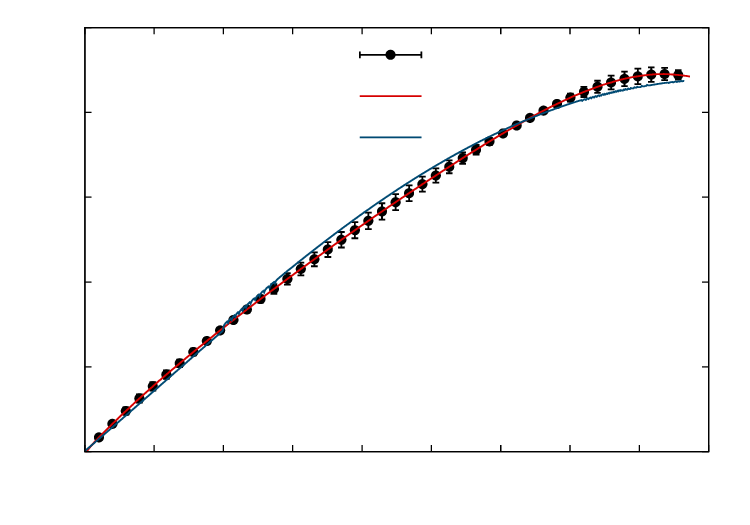}}%
    \gplfronttext
  \end{picture}%
\endgroup

%% file: tension/crack2VST.tex
\begingroup
\large
  \makeatletter
  \providecommand\color[2][]{%
    \GenericError{(gnuplot) \space\space\space\@spaces}{%
      Package color not loaded in conjunction with
      terminal option `colourtext'%
    }{See the gnuplot documentation for explanation.%
    }{Either use 'blacktext' in gnuplot or load the package
      color.sty in LaTeX.}%
    \renewcommand\color[2][]{}%
  }%
  \providecommand\includegraphics[2][]{%
    \GenericError{(gnuplot) \space\space\space\@spaces}{%
      Package graphicx or graphics not loaded%
    }{See the gnuplot documentation for explanation.%
    }{The gnuplot epslatex terminal needs graphicx.sty or graphics.sty.}%
    \renewcommand\includegraphics[2][]{}%
  }%
  \providecommand\rotatebox[2]{#2}%
  \@ifundefined{ifGPcolor}{%
    \newif\ifGPcolor
    \GPcolortrue
  }{}%
  \@ifundefined{ifGPblacktext}{%
    \newif\ifGPblacktext
    \GPblacktextfalse
  }{}%
  \let\gplgaddtomacro\g@addto@macro
  \gdef\gplbacktext{}%
  \gdef\gplfronttext{}%
  \makeatother
  \ifGPblacktext
    \def\colorrgb#1{}%
    \def\colorgray#1{}%
  \else
    \ifGPcolor
      \def\colorrgb#1{\color[rgb]{#1}}%
      \def\colorgray#1{\color[gray]{#1}}%
      \expandafter\def\csname LTw\endcsname{\color{white}}%
      \expandafter\def\csname LTb\endcsname{\color{black}}%
      \expandafter\def\csname LTa\endcsname{\color{black}}%
      \expandafter\def\csname LT0\endcsname{\color[rgb]{1,0,0}}%
      \expandafter\def\csname LT1\endcsname{\color[rgb]{0,1,0}}%
      \expandafter\def\csname LT2\endcsname{\color[rgb]{0,0,1}}%
      \expandafter\def\csname LT3\endcsname{\color[rgb]{1,0,1}}%
      \expandafter\def\csname LT4\endcsname{\color[rgb]{0,1,1}}%
      \expandafter\def\csname LT5\endcsname{\color[rgb]{1,1,0}}%
      \expandafter\def\csname LT6\endcsname{\color[rgb]{0,0,0}}%
      \expandafter\def\csname LT7\endcsname{\color[rgb]{1,0.3,0}}%
      \expandafter\def\csname LT8\endcsname{\color[rgb]{0.5,0.5,0.5}}%
    \else
      \def\colorrgb#1{\color{black}}%
      \def\colorgray#1{\color[gray]{#1}}%
      \expandafter\def\csname LTw\endcsname{\color{white}}%
      \expandafter\def\csname LTb\endcsname{\color{black}}%
      \expandafter\def\csname LTa\endcsname{\color{black}}%
      \expandafter\def\csname LT0\endcsname{\color{black}}%
      \expandafter\def\csname LT1\endcsname{\color{black}}%
      \expandafter\def\csname LT2\endcsname{\color{black}}%
      \expandafter\def\csname LT3\endcsname{\color{black}}%
      \expandafter\def\csname LT4\endcsname{\color{black}}%
      \expandafter\def\csname LT5\endcsname{\color{black}}%
      \expandafter\def\csname LT6\endcsname{\color{black}}%
      \expandafter\def\csname LT7\endcsname{\color{black}}%
      \expandafter\def\csname LT8\endcsname{\color{black}}%
    \fi
  \fi
  \setlength{\unitlength}{0.0500bp}%
  \begin{picture}(7200.00,5040.00)%
    \gplgaddtomacro\gplbacktext{%
      \csname LTb\endcsname%
      \put(682,704){\makebox(0,0)[r]{\strut{} 0}}%
      \put(682,1518){\makebox(0,0)[r]{\strut{} 5}}%
      \put(682,2332){\makebox(0,0)[r]{\strut{} 10}}%
      \put(682,3147){\makebox(0,0)[r]{\strut{} 15}}%
      \put(682,3961){\makebox(0,0)[r]{\strut{} 20}}%
      \put(682,4775){\makebox(0,0)[r]{\strut{} 25}}%
      \put(814,484){\makebox(0,0){\strut{} 0}}%
      \put(1563,484){\makebox(0,0){\strut{} 1}}%
      \put(2311,484){\makebox(0,0){\strut{} 2}}%
      \put(3060,484){\makebox(0,0){\strut{} 3}}%
      \put(3809,484){\makebox(0,0){\strut{} 4}}%
      \put(4557,484){\makebox(0,0){\strut{} 5}}%
      \put(5306,484){\makebox(0,0){\strut{} 6}}%
      \put(6054,484){\makebox(0,0){\strut{} 7}}%
      \put(6803,484){\makebox(0,0){\strut{} 8}}%
      \put(176,2739){\rotatebox{-270}{\makebox(0,0){\strut{}Force $\left[\rm N\right]$}}}%
      \put(4072,154){\makebox(0,0){\strut{}Displacement $\left[\rm mm\right]$}}%
    }%
    \gplgaddtomacro\gplfronttext{%
      \csname LTb\endcsname%
      \put(3322,4514){\makebox(0,0)[r]{\strut{}Standard deviation}}%
      \csname LTb\endcsname%
      \put(3322,4118){\makebox(0,0)[r]{\strut{}Experiment}}%
      \csname LTb\endcsname%
      \put(3322,3722){\makebox(0,0)[r]{\strut{}Simulation}}%
    }%
    \gplbacktext
    \put(0,0){\includegraphics{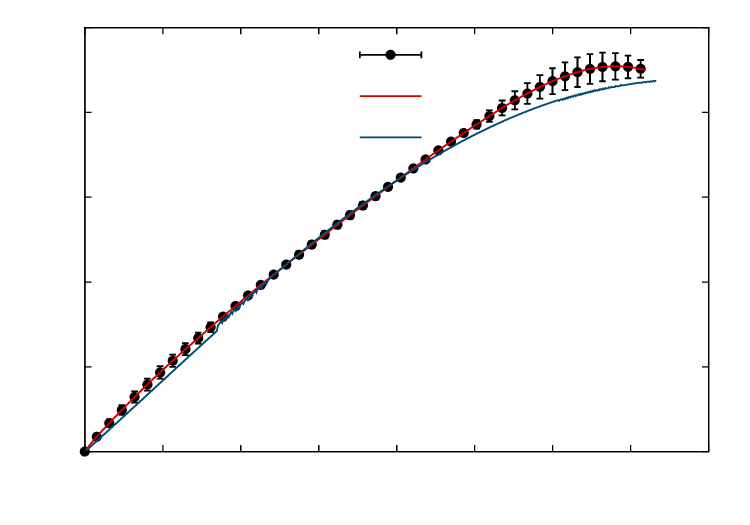}}%
    \gplfronttext
  \end{picture}%
\endgroup